\documentclass [preprint,amsmath,amssymb,showkeys,floatfix]{revtex4}

\usepackage{color}
\usepackage{epic}
\usepackage{eepic}
\usepackage{graphicx,amsmath,amssymb}

\begin{document}

\author{Ashot S. Gevorkyan}
\email{g_ashot@sci.am} \affiliation{Institute for Informatics and
Automation Problems, NAS of Armenia, P. Sevak St. 1, Yerevan
375014,} \affiliation{Institute of Applied Problems in Physics,
NAS Armenia, Nersisian St. 25,
 Yerevan 375014, }

\author{Gabriel G.\ Balint-Kurti}
\email{Gabriel.Balint-Kurti@Bristol.ac.uk} \affiliation {Center
for Computational Chemistry, School of Chemistry, University of
Bristol, Bristol BS8 1TS, UK}

\author{Gunnar Nyman}
\email{nyman@chem.gu.se} \affiliation{G\"{o}teborg University,
Department of Chemistry, SE-412 96, G\"{o}teborg, Sweden}

\title{ A New Approach To The Evaluation Of The
$S$-Matrix In  Atom-Diatom Quantum Reactive Scattering Theory}

\date{\today}

\begin{abstract}
A new approach is described to the evaluation of the
$\textbf{\emph{S}}$-matrix in three-dimensional atom-diatom
reactive quantum scattering theory. The theory is developed based
on natural collision coordinates where progress along the reaction
coordinate can be viewed as fulfilling the same role as time in a
time-dependent formulation. By writing the full wavefunction in
coupled-channel form it is proved that the $3D$ quantum reactive
scattering problem can be treated in the same way as an inelastic
single-arrangement problem. In particularly, two types of
coupled-channel representations, which are reduced to two
different systems of coupled first order ordinary differential
equations describing the inelastic scattering, are used. The first
system of coupled differential equations is constructed on a set
of points (grid) of the coordinate reaction curve after solution
of many $1D$ Schr\"odinger problems in the directions normal to
the reaction coordinate. The second expression for inelastic
scattering is found using exactly solvable nonstationary $1D$
Schr\"odinger equation (etalon equation method), which is
introduced for describing the localization properties of the full
wavefunction along the curve of coordinate reaction. In this case
we avoid a large amount of computation involved in solving the
$1D$ Schr\"odinger problem along the reaction coordinate by using
a slightly difficult initial conditions for the inelastic
scattering equations. In both cases by solving the system of
coupled first order ordinary differential equations, the full
wavefunction and all $\textbf{\emph{S}}$-matrix elements are
obtained simultaneously without further calculations. Our analysis
shows that the methods we have developed constitute the simplest
algorithms for computing the reactive scattering
$\textbf{\emph{S}}$-matrices.
\end{abstract}


\maketitle

\section{Introduction}

Accurate quantum dynamics approaches have made an immense impact
on the theory of elementary atom-molecule collisions
\cite{Schatz76,Kuppermann81,Manolopoulos}. Recent advances in
accurate quantum mechanical calculations on simple few atom
systems have demonstrated the possibility of obtaining detailed
information concerning chemical reactions from first principles
\cite{Launay02,AJCP01,GBKSLLHYJCP00}. Tremendous progress has
occurred in developing and applying both time-independent and
time-dependent quantum dynamics approaches the last few years.

At present there are three common approaches used in the quantum
reactive scattering field:
\begin{enumerate}
\item The hyperspherical coordinate methods \cite{Delves,Schatz},
\item The variational methods based on the simultaneous use of
mass-scaled Jacobi coordinates in each of the chemical
arrangements involved \cite{Miller,BaerKouri,MBaer} and \item
Wavepacket methods \cite{Kosloff,Kurti} (many references can be
found in the review  \cite{Nyman}).
\end{enumerate}

Both time-dependent and time-independent approaches have been
successfully applied to three- and four-atom reactions. For
detailed information at low energy the time-independent
hyperspherical coordinate approach has proven the most useful
\cite{Launay02,Manolopoulos2}. The time-dependent wavepacket
method has also proved to be very useful, particularly for
photodissociation and laser-molecule interaction problems. To go
beyond four-atom systems, the Multi-Configurational Time-Dependent
Hartree method has been applied in full-dimensionality to
calculate thermal rate constants for the H + CH$_4$  \cite{MZS}
and O + CH$_4$ \cite{HNyman} reactions.

Despite the successful applications to polyatomic systems
mentioned above, general applications still appear to be extremely
difficult. Quantum mechanics is nonlocal in character and
numerical efforts to solve the Schr\"odinger equation increase
exponentially with the number of degrees of freedom. State
resolved accurate quantum dynamical calculations are presently
limited to at most seven-dimensional problems \cite{MZS}. Thus,
the development of new conceptual approaches permitting improved
computational algorithms for accurate  quantum simulations is of
great interest.

Earlier \emph{natural collision coordinates} NCC  allowing an
analytical investigation of chemical reactivity in collinear
collisions were  introduced by Marcus \cite{Marcus} and further
explored by Light \cite{Light,WLAJCP76}. Light and coworkers
\cite{LightW,SWL} generalized the NCC approach to
three-dimensional atom-diatom reactive scattering
 and on this basis they studied
H + H$_2$ system and its isotopomers in $3D$ for  total angular
momentum $J=0$ \cite{LightAl}. Related developments have been the
formulation and later use of the reaction-path hamiltonian
formalism by Miller, Handy and Adams \cite{MHAJCP80} and the
reaction path and reaction volume approaches advocated by Billing
\cite{BMP96,KBJCP97,CBPCCP99,BCP02}.

Due to difficulties in applying the NCC approach, attention turned
to other coordinate systems. Nevertheless, it seems to us that the
primary idea of NCC, namely to simplify $3D$ quantum reactive
scattering calculations, can be realized. Particularly, one of the
authors has shown that for collinear three-body collisions, it is
possible to constructed the $\emph{\textbf{S}}$-matrix exactly, in
simple form using the approach \cite{Ash,Ash1}. Seven reactive
systems, N$_2$ + N, N$_2$ + O, O$_2$ + O, Li + FH, O$_2$ + N etc
have been studied successfully within this framework \cite{Ash2}.

The aim of the current article is to generalize our previous
collinear NCC quantum scattering approach to a full three
dimensional treatment.

\section{ Quantum multi-channel scattering problem in different coordinate systems}

The reactive scattering process in a three-body system can be
described by the scheme:

$$
 A+(BC)_n \longrightarrow
 \begin{cases}
A+(BC)_m ,\\
(AB)_m + C, \\
(AC)_{p}+B,\\
A+B+C, \\
\quad\quad\qquad(ABC)^*\longrightarrow
\begin{cases}
A+(BC)_m ,\\
(AB)_m + C ,\\
(AC)_{p}+B,\\
 A+B+C,
\end{cases}
 \end{cases}
 $$
$$
Sch.\,\, 1.
$$
 where $A,B,$ and $C$ are atoms, $n$ and $m, p$ characterize
quantum numbers of diatomic states corresponding to initial $(in)$
and final $(out)$ scattering channels,  $(ABC)^*$ denotes an
activated complex which may or may not be present as an
intermediary species.

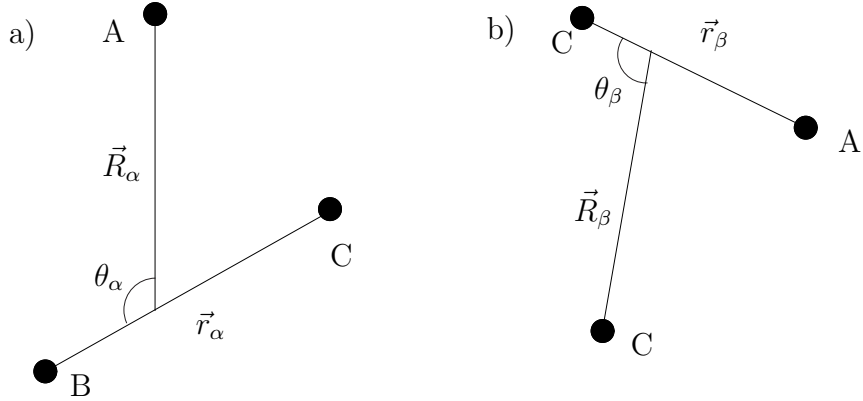
\begin{figure}
\begin{center}
\setlength{\unitlength}{0.00047489in}
\begingroup\makeatletter\ifx\SetFigFont\undefined%
\gdef\SetFigFont#1#2#3#4#5{%
  \reset@font\fontsize{#1}{#2pt}%
  \fontfamily{#3}\fontseries{#4}\fontshape{#5}%
  \selectfont}%
\fi\endgroup%
{\renewcommand{\dashlinestretch}{30}
\begin{picture}(9426,4380)(0,-10)
\put(1627.500,952.500){\arc{705.160}{2.7255}{4.6911}}
\put(7053.750,3780.000){\arc{630.402}{1.5351}{3.7494}}
\put(1620,4230){\blacken\ellipse{254}{254}}
\put(1620,4230){\ellipse{254}{254}}
\put(3555,2070){\blacken\ellipse{254}{254}}
\put(3555,2070){\ellipse{254}{254}}
\put(405,270){\blacken\ellipse{254}{254}}
\put(405,270){\ellipse{254}{254}}
\put(6570,720){\blacken\ellipse{254}{254}}
\put(6570,720){\ellipse{254}{254}}
\put(8820,2970){\blacken\ellipse{254}{254}}
\put(8820,2970){\ellipse{254}{254}}
\put(6345,4185){\blacken\ellipse{254}{254}}
\put(6345,4185){\ellipse{254}{254}}
\path(1620,4230)(1620,945)
\path(405,270)(3600,2070)
\path(6345,4185)(8820,2970)
\path(7110,3825)(6570,720)
\put(1035,3915){\makebox(0,0)[lb]{{\SetFigFont{12}{14.0}{\rmdefault}{\mddefault}{\updefault}A}}}
\put(1035,2385){\makebox(0,0)[lb]{{\SetFigFont{12}{14.0}{\rmdefault}{\mddefault}{\updefault}$\vec{R}_\alpha$}}}
\put(945,1170){\makebox(0,0)[lb]{{\SetFigFont{12}{14.0}{\rmdefault}{\mddefault}{\updefault}$\theta_\alpha$}}}
\put(2070,630){\makebox(0,0)[lb]{{\SetFigFont{12}{14.0}{\rmdefault}{\mddefault}{\updefault}$\vec{r}_\alpha$}}}
\put(675,0){\makebox(0,0)[lb]{{\SetFigFont{12}{14.0}{\rmdefault}{\mddefault}{\updefault}B}}}
\put(3555,1440){\makebox(0,0)[lb]{{\SetFigFont{12}{14.0}{\rmdefault}{\mddefault}{\updefault}C}}}
\put(6000,3800){\makebox(0,0)[lb]{{\SetFigFont{12}{14.0}{\rmdefault}{\mddefault}{\updefault}C}}}
\put(9180,2700){\makebox(0,0)[lb]{{\SetFigFont{12}{14.0}{\rmdefault}{\mddefault}{\updefault}A}}}
\put(6885,450){\makebox(0,0)[lb]{{\SetFigFont{12}{14.0}{\rmdefault}{\mddefault}{\updefault}C}}}
\put(6255,1845){\makebox(0,0)[lb]{{\SetFigFont{12}{14.0}{\rmdefault}{\mddefault}{\updefault}$\vec{R}_\beta$}}}
\put(6480,3195){\makebox(0,0)[lb]{{\SetFigFont{12}{14.0}{\rmdefault}{\mddefault}{\updefault}$\theta_\beta$}}}
\put(7650,3825){\makebox(0,0)[lb]{{\SetFigFont{12}{14.0}{\rmdefault}{\mddefault}{\updefault}$\vec{r}_\beta$}}}
\put(0,3825){\makebox(0,0)[lb]{{\SetFigFont{12}{14.0}{\rmdefault}{\mddefault}{\updefault}a)}}}
\put(5290,3870){\makebox(0,0)[lb]{{\SetFigFont{12}{14.0}{\rmdefault}{\mddefault}{\updefault}b)}}}
\end{picture}
}
\end{center}
\caption{\emph{ Jacobi coordinates for the reaction
$A+(BC)_n\rightarrow (AB)_m+C$ corresponding to the reactant (a)
and to the product (b) channels.}} \label{fig:fig1}
\end{figure}

\subsection{The coordinate systems}

We denote the masses of the particles by $m_A,m_B$ and $m_C$ and
the coordinates of the particles are denoted by the column vectors
${\bf{r_{A}}}, {\bf{r_{B}}}$ and ${\bf{r_{C}}}$ describing their
positions relative to an origin fixed in the laboratory system.
The initial $9D$ problem can be reduced to a $6D$ one by
elimination of the center-of-mass coordinates.

 Below we consider the
reaction $A+(BC)_n\rightarrow(ABC)^\ast\rightarrow(AB)_m+C$, i.e.
there are two open arrangements.

For this reaction reactant and product Jacobi coordinates are
illustrated in Fig.1. The following equations define the Jacobi
coordinates $\bf{R}_\alpha$ and $\bf{r}_\alpha$ (note that index
$\alpha$ corresponds to $(in)$ or reactant asymptotic channel):
\begin{eqnarray}
&&{\bf{R}}_\alpha={\bf{r}}_A-\frac{m_B{\bf{r_B}}+m_C{\bf{r_C}}}{m_B+m_C},
\qquad {\bf{r}}_\alpha= {\bf{r_C}}-{\bf{r_B}},\qquad \qquad
\nonumber \\
&&\Bigl({\bf{R}}_\alpha\equiv{\bf{R}}_\alpha(R_x,R_y,R_z),\,\,\,
{\bf{r}}_\alpha\equiv{\bf{r}}_\alpha(r_x,r_y,r_z)\Bigr).
\label{01}
\end{eqnarray}

We now apply the Delves-Smith scale transformation
\cite{Delves,Smith,MBaer} to obtain the new scaled coordinates
$\bf{q}_0$ and $\bf{q}_1$;
\begin{eqnarray}
\label{02}
{\bf{q}_0}\equiv{\bf{q}_{0\alpha}}=\lambda\,{\bf{R}}_\alpha,\qquad
{\bf{q}_1}\equiv{\bf{q}_{1\alpha}}=\lambda^{-1}\,{\bf{r}}_\alpha.
\qquad
\end{eqnarray}
where
\begin{eqnarray}
\label{03}
 \lambda=\Bigl[\frac{m_A}{\mu}\Bigl(1-\frac{m_A}{M}\Bigr)\Bigr]^{1/2},
 \qquad
\mu=\Bigl[\frac{m_Am_Bm_C}{M}\Bigr]^{1/2},\qquad M=m_A+m_B+m_C.
\end{eqnarray}

In these coordinates the Hamiltonian of a three-body system takes
the following diagonal form:
\begin{eqnarray}
\label{04}
H\bigl({\bf{q}};{\bf{P_{{\bf{q}}}}}\bigr)=\frac{1}{2\mu}
{\bf{P}}_{\bf{q}}^2+
V\bigl(q_0,q_1,\theta\bigr),\qquad\qquad\qquad
\end{eqnarray}
where
$$
{\bf{q}}=\bigl({\bf{q}_0},{\bf{q}_1})=\{q_k\}, \quad
k=0,...,5;\quad
\{q_k\}=\bigl(q_0=|{\bf{q_0}}|,\,\,q_1=|{\bf{q_1}}|,\,\,q_2,\,\,
q_3,\,\,q_4,\,\,q_5\bigr).
$$
Note that here and in the following we omit the $\alpha$ channel
index for simplicity. In Eq. (\ref{04}) $\mu$ is the reduced mass
defined in Eq. (\ref{03}),  ${\bf{P}_{\bf{q}}}$ is the moment
vector conjugate to  ${\bf{q}}$, in the body-fixed system and
$\theta$ is the angle between vectors ${\bf{\bf{q_0}}}$ and
${\bf{q}_1}$ (see FIG. 1 and FIG. 2). An alternative coordinate
system $(q_0,q_1,q_2)$ is illustrated in FIG. 2. Either of the
coordinate systems $(q_0,q_1,q_2)$ and $(q_0,q_1,\theta)$ can be
used to describe the motion of body-fixed system in the plane of
the vectors $({\bf{q}_0},{\bf{q}_1})$ and they will be called
intrinsic coordinates. The interaction potential between all atoms
depends only on the intrinsic coordinates.

  The coordinates $(q_3,q_4,q_5)$ are taken to be the Euler
angles $\Omega(\equiv\zeta_1,\zeta_2,\zeta_3)$, which orient the
three-body system in the space-fixed frame (see for example refs.
\cite{Kurti}, \cite{Edmonds} and \cite{Zare}).

The optimal coordinate systems for reactants and products are
different \cite{Miller,PackPark}. This fact creates certain
mathematical and computational complexities in the investigation
of the reactive scattering problem. One way to proceed is to turn
to a special type of curvilinear coordinates which continuously
leads from reactant channel to product channel. The  possibility
of introducing the coordinate system which is simultaneously
suitable for description of $(in)$ and $(out)$ asymptotic states
 was first time discussed in the work \cite{Hof}. At last corrected
version for collinear collision was presented in the work
\cite{Marcus} and named \emph{natural collision coordinate} (NCC)
system.

In order to define the NCC  appropriately, we connect the $(in)$
and $(out)$ asymptotic channels by a smooth curve $\Im_{if}$. The
curve depends on two parameters, one of which can be chosen
arbitrary and the other is system dependent. As progress along the
curve represent the passage from reactant to product during this
it will be called \emph{reaction coordinate } \cite{Marcus,Light}.
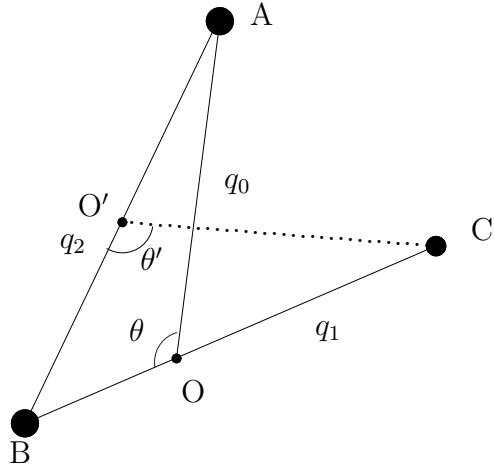
\begin{figure}
\begin{center}
\setlength{\unitlength}{0.00050489in}
\begingroup\makeatletter\ifx\SetFigFont\undefined%
\gdef\SetFigFont#1#2#3#4#5{%
  \reset@font\fontsize{#1}{#2pt}%
  \fontfamily{#3}\fontseries{#4}\fontshape{#5}%
  \selectfont}%
\fi\endgroup%
{\renewcommand{\dashlinestretch}{30}
\begin{picture}(4534,4785)(0,-10)
\put(1813.677,1044.265){\arc{636.671}{2.9703}{4.4301}}
\put(2175,4590){\blacken\ellipse{284}{284}}
\put(2175,4590){\ellipse{284}{284}}
\put(1162,2497){\arc{636.671}{0.15}{2.1}}
\put(1162,2497){\blacken\ellipse{90}{90}}
\put(1162,2497){\ellipse{90}{90}}
\dottedline[.]{100}(1162,2497)(4425,2250)
\put(4425,2250){\blacken\ellipse{202}{202}}
\put(4425,2250){\ellipse{202}{202}}
\put(150,405){\blacken\ellipse{284}{284}}
\put(150,405){\ellipse{284}{284}}
\put(1725,1080){\blacken\ellipse{90}{90}}
\put(1725,1080){\ellipse{90}{90}}
\path(2175,4590)(1725,1080)
\path(150,405)(4425,2250)
\path(150,405)(2175,4635)
%\path(150,405)(2175,4635)
\put(2490,4545){\makebox(0,0)[lb]{{\SetFigFont{12}{14.0}{\rmdefault}{\mddefault}{\updefault}A}}}
\put(700,2560){\makebox(0,0)[lb]{{\SetFigFont{12}{14.0}{\rmdefault}{\mddefault}{\updefault}O$'$}}}
\put(1350,2000){\makebox(0,0)[lb]{{\SetFigFont{12}{14.0}{\rmdefault}{\mddefault}{\updefault}$\theta'$}}}
\put(4785,2310){\makebox(0,0)[lb]{{\SetFigFont{12}{14.0}{\rmdefault}{\mddefault}{\updefault}C}}}
\put(-15,-13){\makebox(0,0)[lb]{{\SetFigFont{12}{14.0}{\rmdefault}{\mddefault}{\updefault}B}}}
\put(1770,630){\makebox(0,0)[lb]{{\SetFigFont{12}{14.0}{\rmdefault}{\mddefault}{\updefault}O}}}
\put(3165,1260){\makebox(0,0)[lb]{{\SetFigFont{12}{14.0}{\rmdefault}{\mddefault}{\updefault}$q_{1}$}}}
\put(2220,2790){\makebox(0,0)[lb]{{\SetFigFont{12}{14.0}{\rmdefault}{\mddefault}{\updefault}$q_{0}$}}}
\put(510,2160){\makebox(0,0)[lb]{{\SetFigFont{12}{14.0}{\rmdefault}{\mddefault}{\updefault}$q_{2}$}}}
\put(1230,1260){\makebox(0,0)[lb]{{\SetFigFont{12}{14.0}{\rmdefault}{\mddefault}{\updefault}$\theta$}}}
\end{picture}
}
\end{center}
\caption{\emph{The three-body system in  mass-scaled Jacobi
coordinates. The coordinate $q_0$ shows the distance between
particle $A$ and the center of mass $(BC)$ pair $O$, the distance
between particles $B$ and $C$ is $q_1$ and $q_2$ correspondingly
between $A$ and $C$, \, \,$\theta$ is a Jacobi scattering angle.
The point $O^{\,'}$ is a center of mass diatom $(AB)$ and
$\theta^{\,'}$ is correspondingly Jacobi scattering angle which is
used in $(out)$ channel.}}
\end{figure}

 The curve $\Im_{if}$, which connects $(in)$ and
$(out)$ asymptotic channels can be determined in the plane
($q_0,q_1,\theta=0$), by simple formula:
\begin{equation}
q_{0}^c=
\frac{a}{q_1^c-q_{eq}^-}+b\bigl(q_1^c-q_{eq}^-\bigr)+q^+_{eq},\qquad
q_{eq}^-<q_1^c<+\infty, \label{05}
\end{equation}
where $a$ and $b$  are constants.

Note that  the collinear collision configuration $A+(BC)$
describes on the plane ($q_0,q_1,\theta=0$) while the collinear
collision $A+(CB)$ describes on the plane ($q_0,q_1,\theta=\pi$).
The motivation of choice of curve $\Im_{if}$ for later definition
of $3D$ NCC system is given in the Appendix A. In Eq. (\ref{05})
$q^-_{eq}$ and $q^+_{eq}$ are mass-scaled equilibrium bond lengths
of molecules in the $(in)$ and $(out)$ channels correspondingly.
Note that $a$ is an arbitrary constant, which is usually chosen to
make the curve pass close to the saddle point of the reaction. The
constant $b$ depends on the mass of particles and is given by
expression $b=cot\varphi=\bigl[{m_Am_C}/{m_BM}\bigr]^{1/2}$. The
superscript $c$ over $q_0$ and $q_1$ underlines the fact that the
point $(q_0^c,q_1^c)$ lies on the curve.

The limit $q_1^c\rightarrow q_{eq}^-,\,\,q_0^c\rightarrow{\infty}$
corresponds to the reactant region or $(in)$ channel while the
limit $q_1^c\rightarrow\infty,\,\, q_0^c\rightarrow\infty$ on the
curve $\Im_{if}$ corresponds to the product region or $(out)$
channel, where $q_0^c=bq_1^c+q^+_{eq}$. It is easy to show that
the mass-scaled distance $q_{2}$ between $A$ and $B$ particles
connected with the curve $\Im_{if}$ is described by expression
$q_{2}^c=q_{2}\bigl|_{\Im_{if}}=\sqrt{(q_{0}^c)^2-2bq_{0}^c
q_{1}^c\cos\theta+b^2(q_{1}^c)^2}$, which in the $(in)$ channel is
equal $\bigl(q_{2}^c-q_0^c\bigr)\bigl|_{q_{1}^c\to
q_{eq}^-}=-bq_{eq}^-\cos\theta$. In other words it is denote that
in the $(in)$  channel curve $\Im_{if}$ is coming to the
equilibrium
 distance $q_{eq}^-$ of  $(BC)$ diatom. In the $(out)$ channel the
 curve $\Im_{if}$ approaches the equilibrium distance $q^+_{eq}$ of diatom
$(AB)$ (see Appendix B). Note that the Jacobi scattering angle
$\theta$ which defined in the reactant (or $(in)$) coordinates
system  in the product (or $(out)$) channel obviously must be
limited to $0$ inasmuch as the two sides $\overline{BO}=bq_1$ and
$\overline{OA}=q_0$ of $\widehat{ABO}$ triangle are limited to
infinity while the third side is limited to finite quantity
$\overline{AB}= q_2\to q_{eq}^+$. Remaind that it doesn't mean
that the rotation of new-formed diatom $(AB)$ stops. The rotation
of diatom $(AB)$ is described by angle $\theta'$ (see  FIG. 2)
which in the reactant region is limited to $0$ but beginning from
strong interaction region up to $(out)$ asymptotic channel
obviously can have any value.

The coordinate $u$  describes the translational motion of
three-body system between reactant and product channels and is
changed along the curve $\Im_{if}$ measured from an initial point
$u_0$. It particularly can be determined by equation:
\begin{equation}
u= u_0-\frac{a}{q_1^c-q_{eq}^-}+b(q_1^c-q_{eq}^-\bigr).\label{06}
\end{equation}
 The signed distance
from the curve $\Im_{if}$ in the plane ($q_0,q_1,\theta=0$) is
given by $v$. We can organize the one-to-one mapping between
coordinate systems $(q_0,q_1,\theta)$ and $(u,v,\theta)$ in only
some subspace of internal 3D configuration space. It should be
emphasized that above mentioned subspace must include  a part of
3D space, in which the probability current of reactive scattering
process is localized. This condition can be fulfilled if the
coordinate reaction curve $\Im_{if}$ is correctly determined for a
configuration of reactive collision with the minimal energy
surface (see FIG. 3). It is obvious that in this case the
condition of one-to-one mapping between coordinate systems for any
configuration of reactive collision can be automatically
satisfied.

After satisfying aforementioned condition it is easy to write
transformations between set of coordinates $(q_0,q_1)$ and $(u,v)$
\cite{Light2}:
\begin{eqnarray}
\label{07}
q_0(u,v)=q_0^c(u)-v\sin\phi(u),\nonumber\\
q_1(u,v)=q_1^c(u)+v\cos\phi(u),
\end{eqnarray}
where the angle $\phi(u)$ is determined from the requirement that
the coordinate system $(u,v)$ should be orthogonal (see Appendix
C):
\begin{figure}
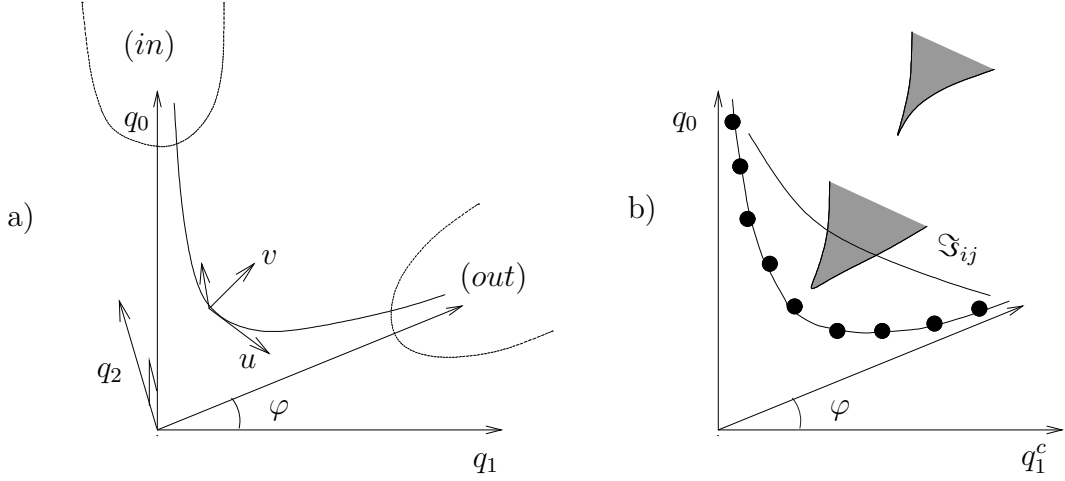

\begin{center}
\input fig3.eepic
\end{center}
\caption{a) \emph{Intrinsic 3D Jacobi and natural collision
coordinate (NCC) systems. The angle $\varphi$  is defined in the
text. b) The reaction path (cycled curve) is passing through the
minimums of potential energy while the reaction coordinate
$\Im_{if}$ can be an arbitrary smooth curve connecting $(in)$ and
$(out)$ asymptotic channels. The lower shaded area is
self-crossing region for NCC system associated to the reaction
path curve while the upper one self-crossing region for NCC system
is associated to the curve $\Im_{if}$. }} \label{fig:fig3}
\end{figure}
\begin{equation}
\frac{dq_0^c}{dq_1^c}\biggl|_{(u,v=0)}=\cot\phi(u),\qquad
\lim_{u\rightarrow+\infty}\cot\phi(u)=\cot\varphi. \label{08}
\end{equation}
In the NCC system the  motions of the body system are locally
factorized  into translational $\bigl(u$, infinite
extension$\bigr)$, vibrational $\bigl(v$, finite$\bigr)$ and
intrinsic rotational ($\theta$, finite) motions. Note that the
coordinate $u$ is perpendicular to the curve $v$ and independent
from the angle $\theta$. These properties ensure that the full
wavefunction can be conditionally (locally) factorized.

So, we defined NCC system for investigation of $3D$ reactive
quantum scattering which for using has only one limitation related
with the region of self-crossing of coordinate lines. In other
words it is necessary to choose  the reaction coordinate
$\Im_{if}$ in such a way in order to the probability current of
quantum reactive scattering process in the region of self-crossing
of coordinate line  was absent.

\section{Quantum reactive scattering in the  NCC
system}

\subsection{Equation of motion of the three-body system}

The overall stationary wavefunction of three-body system  can be
written as:
\begin{equation}
\Psi^{JM}(q_0,q_1,\theta,\Omega)=\frac{1}{q_0q_1}\sum_{K}
\overline{\Phi}^{J}_{K}(q_0,q_1,\theta)D^J_{KM}(\Omega),
 \label{09}
\end{equation}
 where $J$ is the total angular momentum, $M$ and $K$ are its
space-fixed and  body-fixed $z$-components respectively. The
summation over $K$ ranges from $-J$ to $+J$ in unit steps and
$D^J_{KM}$ is the Wigner $D$-matrix \cite{Edmonds,Zare}.

 After separation of the external rotations, the action of the
hamiltonian on the wavefunction is given by the formula
\cite{Kurti,WLAJCP76}:
\begin{eqnarray}
\hat{H}\overline{\Phi}^{J}_{K}(q_0,q_1,\theta)&=&
-\frac{\hbar^2}{2\mu}\biggl\{ \frac{\partial^2}{\partial{q_0^2}}+
\frac{\partial^2}{\partial{q_1^2}}\biggr\} \,
\overline{\Phi}^{J}_{K}(q_0,q_1,\theta)
\qquad\qquad\qquad\qquad\qquad\qquad\,\,\qquad
\nonumber\\
&&- \, \frac{\hbar^2}{2 \mu}\biggl( \frac{q_0^2+q_1^2}{q_0^2q^2_1}
\biggr)
\biggl\{\frac{1}{\sin\theta}\frac{\partial}{\partial{\theta}}\sin\theta
\frac{\partial}{\partial{\theta}}
-\frac{K^2}{\sin^2\theta}\biggr\} \,
\overline{\Phi}^{J}_{K}(q_0,q_1,\theta)\qquad\quad\,\,\,\,
\nonumber\\
&&+ \,
V(R,r,\theta)\overline{\Phi}^{J}_{K}(q_0,q_1,\theta)+\frac{\hbar^2}{2
\mu{q_0^2}} \Bigl(J(J+1)-2K^2\Bigr) \,
\overline{\Phi}^{J}_{K}(q_0,q_1,\theta)\qquad
\nonumber\\
&&- \, \frac{\hbar^2}{2
\mu{q_0^2}}\,c^+_{JK}\biggl\{\frac{\partial}{\partial{\theta}}
-(K+1)\cot\theta\biggr\} \,
\overline{\Phi}^{J}_{(K+1)}(q_0,q_1,\theta)
\nonumber\\
&& + \, \frac{\hbar^2}{2
\mu{q_0^2}}\,c^-_{JK}\biggl\{\frac{\partial}{\partial{\theta}}
+(K-1)\cot\theta\biggr\} \,
\overline{\Phi}^{J}_{(K-1)}(q_0,q_1,\theta)
\nonumber \\
&=&  E \, \overline{\Phi}^{J}_{K}(q_0,q_1,\theta),
 \label{10}
\end{eqnarray}
 where
\begin{equation}
c^{\pm}_{JK}=\bigl[J(J+1)-K(K\pm1)\bigr]^{1/2}.
 \label{11}
\end{equation}

After the coordinate transformation
$(q_0,q_1,\theta)\rightarrow(u,v,\theta)$  in equation (\ref{10})
we find:

\begin{eqnarray}
\hat{H}\overline{\Phi}^{J}_{K}(u,v,\theta)&=&-\frac{\hbar^2}{2\mu}\frac{1}{\eta}
\biggl\{\frac{\partial}{\partial{u}}\frac{1}{\eta}
\frac{\partial}{\partial{u}}+\frac{\partial}{\partial{v}} \eta\,
\frac{\partial}{\partial{v}}\biggr\}
\overline{\Phi}^{J}_{K}(u,v,\theta)-\frac{\,\hbar^2}{2
\mu}\frac{1}{\,q^2(u,v)}\times
\nonumber\\
&&
\times\biggl\{\frac{1}{\sin\theta}\frac{\partial}{\partial{\theta}}\sin\theta
\frac{\partial}{\partial{\theta}}
-\frac{K^2}{\sin^2\theta}\biggr\}\overline{\Phi}^{J}_{K}(u,v,\theta)
+U(u,v,\theta)\overline{\Phi}^{J}_{K}(u,v,\theta)
\nonumber\\
&& +E_{JK}\overline{\Phi}^{J}_{K}(u,v,\theta)
+\frac{\hbar^2}{2\mu}\biggl\{H^{+}_{JK}\overline{\Phi}^{J}_{(K+1)}(u,v,\theta)
+H^{-}_{JK}\overline{\Phi}^{J}_{(K-1)}(u,v,\theta)\biggr\}
\nonumber\\
&=& E \, \overline{\Phi}^{J}_{K}(u,v,\theta),
 \label{12}
\end{eqnarray}
 where
$$
E_{JK}(u,v)=\frac{\hbar^2}{2 \mu q_0^2(u,v)}
\Bigl(J(J+1)-2K^2\Bigr),\quad H^{\pm}_{JK} =
\frac{c^{\pm}_{JK}}{q_0^2(u,v)}\,
\biggl\{\mp\frac{\partial}{\partial{\theta}}
+\,(K\pm\,1)\cot\theta\biggr\},
$$
\begin{eqnarray}
q^2(u,v)=\frac{q_0^2(u,v)q_1^2(u,v)}{q_0^2(u,v)+q_1^2(u,v)}, \quad
V(R,r,\theta)\equiv\,U(u,v,\theta),\quad
 \overline{\Phi}^{J}_{K}(q_0,q_1,\theta)\equiv
\overline{\Phi}^{J}_{K}(u,v,\theta).\quad
 \label{13}
\end{eqnarray}

 In expression (\ref{12}) the Lam\'{e} coefficient
$\eta(u,v)$ has a form (Appendix C):

\begin{equation}
 \eta(u,v)=\bigl[1+K(u)v\bigr]\frac{ds}{du},
 \label{14}
 \end{equation}
 where $K(u)$ is described curvature of the
reaction coordinate $\Im_{ij}$ in the continues point $u$,
correspondingly $s$ the length along the $\Im_{ij}$:
\begin{equation}
K(u)= \frac{2a}{\bigl(q_1^c-q_{eq}^-\bigr)^3}\biggl\{1+
\Bigl[b-a/{\bigl(q_1^c-q_{eq}^-\bigr)^2}
 \Bigr]^2\biggr\}^{-3/2},
 \label{15}
\end{equation}
and
\begin{equation}
\frac{ds}{du}=\biggl\{1+\Bigl[b-
a/{\bigl(q_1^c-q_{eq}^-\bigr)^2}\Bigr]^2
 \biggr\}^{1/2}\biggl\{b+a/{\bigl(q_1^c-q_{eq}^-\bigr)^2}\biggr\}^{-1}.
 \label{16}
\end{equation}

Finally we transform Eq. (\ref{12}) by setting
$\overline{\Phi}^J_{K}(u,v,\theta)=\eta^{-1/2}{\Phi}^{J}_{K}(u,v,\theta)$,
whereby we obtain:
\begin{eqnarray}
\biggl\{\biggl[\frac{\partial}{\partial{u}}\frac{1}{\eta^2}
\frac{\partial}{\partial{u}}+
\frac{\partial^2}{\partial{v^2}}\biggr] + \frac{1}{q^2(u,v)}
\biggl[\frac{1}{\sin\theta}\frac{\partial}{\partial{\theta}}\sin\theta
\frac{\partial}{\partial{\theta}}
-\frac{K^2}{\sin^2\theta}\biggr]+\frac{2\mu}{\hbar^2}
\bigl[E-U(u,v,\theta)-\qquad
\nonumber\\
E_{JK}(u,v)-U_{eff}(u,v)\bigr]\biggr\}{\Phi}^{J}_{K}(u,v,\theta)
+\biggl\{H^{+}_{JK}{\Phi}^{J}_{(K+1)}(u,v,\theta)
+H^{-}_{JK}{\Phi}^{J}_{(K-1)}(u,v,\theta)\biggr\}=0,\nonumber\\
 \label{17}
\end{eqnarray}
where the effective potential $U_{eff}(u,v)$ is defined by

\begin{eqnarray}
U_{eff}(u,v)=\frac{1}{4\eta^2}
\biggl(\frac{\partial\eta}{\partial{v}}\biggr)^2-\frac{1}{2\eta^3}
\frac{\partial^2\eta}{\partial{u^2}}+\frac{5}{4\eta^4}
\biggl(\frac{\partial\eta}{\partial{u}}\biggr)^2.
 \label{18}
\end{eqnarray}

Schr\"odinger equation (\ref{17}) is in a form which is suitable
for further investigation  by the coupled channels method. This
will be done in section III. C
 after the $\emph{\textbf{S}}$-matrix has been defined in Section III. B.

\subsection{Definition of ${\bf{\emph{S}}}$-matrix in
 term of reaction coordinate}

In the standard scattering theory the main problem is the
construction of the scattering $\textbf{\emph{S}}$-matrix, holding
the transition amplitudes. Let us discuss the exact representation
for the $\emph{\textbf{S}}$-matrix in terms of overlap between
stationary wavefunctions (see for example \cite{New,GoldWat}). In
the body-fixed NCC system we can write the following general
formal expression:
\begin{equation}
\label{19} \Phi^{(+)J}_{njK}(u_\alpha,v_\alpha,\theta_\alpha)
=\sum_{n'j\,'K'}
\emph{S}^{J}_{n'j\,'K'\,\leftarrow\,nj\,K}\,\Phi^{(-)J}_{n'j\,'K'}
(u_\beta,v_\beta,\theta_\beta),
\end{equation}
where $\Phi^{(+)J}_{njK}(u_\alpha,v_\alpha,\theta_\alpha)$ and
$\Phi^{(-)J}_{n'j\,'K'}(u_\beta,v_\beta,\theta_\beta)$ are the
total stationary wavefunctions in correspondingly NCC systems
which are evaluating from some clear $(in)$ and $(out)$ asymptotic
states, $n$ and $j$ are vibration and rotation quantum numbers,
$\emph{S}_{n'j\,'K'\,\leftarrow\,njK}^{J}$ are an complex matrix
elements.

For the overall wavefunction ${\Phi}^{(+)J}_{njK}(u,v,\theta)$ we
may enforce the following asymptotic behaviors and boundary
conditions:
\begin{eqnarray}
\label{20}
 {\Phi}^{(+)J}_{njK}(u,v,\theta)\,\,\stackrel{\sim}{ _{u
\rightarrow -\infty}}\,\,
{\Phi}_{njK}^{(in)J}(u,v,\theta),\qquad\qquad\qquad\qquad\qquad
\nonumber\\
 {\Phi}^{(+)J}_{njK}(u,v,\theta)\,\, \stackrel{\sim}{ _{u
\rightarrow +\infty}}\,\,
\sum_{n'j\,'K'}S^{J}_{njK\,\leftarrow\,n'j\,'K'}\,(E)\,
{\Phi}_{n'j\,'K'}^{(out)J}(u,v,\theta),
\nonumber\\
\lim_{|v|\rightarrow\infty}{\Phi}^{(+)J}_{njK}(u,v,\theta)
=\lim_{|v|\rightarrow\infty}\frac{\partial}{\partial{v}}
{\Phi}^{(+)J}_{njK}(u,v,\theta)=0.\qquad\qquad\quad\quad
 \end{eqnarray}
where ${\Phi}_{njK}^{(in)J}(u,v,\theta)$ and
${\Phi}_{n'j\,'K'}^{(out)J}(u_\beta,v_\beta,\theta_\beta)$ are an
asymptotic wavefunctions correspondingly in the $(in)$ and $(out)$
channels, $\emph{S}^{J}_{njK\,\leftarrow\,n'j\,'K'}$ are a
reactive $\emph{\textbf{S}}$ -matrix elements.

Similar conditions can be also written for wavefunction
${\Phi}^{(-)J}_{njK} (u,v,\theta)$. For later calculations it is
important to define the behavior of the wavefunction
${\Phi}^{(-)J}_{njK} (u,v,\theta)$ in the $(out)$ asymptotic
channel:
\begin{equation}
\label{21}
 {\Phi}^{(-)J}_{njK} (u_\beta,v_\beta,\theta_\beta) \,\,
\stackrel{\sim}{ _{u \rightarrow +\infty}}\,\,
{\Phi}_{njK}^{(out)J}(u_\beta,v_\beta,\theta_\beta).
\end{equation}
Note that in (\ref{20}) and (\ref{21}) the functions
${\Phi}_{njK}^{(in)J}(u,v,\theta)$ and
${\Phi}_{njK}^{(out)J}(u_\beta,v_\beta,\theta_\beta)$ are
described by expressions:
\begin{eqnarray}
\label{22} {\Phi}_{njK}^{(in)J}(u,v,\theta)=\frac{1}{\sqrt{2\pi}}
\exp\bigl(-ip^{-}_{nj}\,u\bigr)
\Pi_{n(j)}^{(in)}(v)\Theta_{jK}(\theta),
\qquad\qquad\qquad\qquad\qquad\qquad
\nonumber\\
{\Phi}_{njK}^{(out)J}(u_\beta,v_\beta,\theta_\beta)=\frac{1}{\sqrt{2\pi}}
\exp\bigl(ip^{+}_{nj}\,u_\beta\bigr)
\Pi_{n(j)}^{(out)}(v_\beta)\Theta_{jK}(\theta_\beta),\quad
p^{\mp}_{nj}=\frac{1}{\hbar}\sqrt{2\mu{\bigl(E-\epsilon^{\mp}_{n(j)}\bigr)}},
\quad
\end{eqnarray}
where $\epsilon^{-}_{n(j\,)}$ is the vibration-rotational energy
of the initial state of the diatomic, $\epsilon^{+}_{n(j)}$ is
vibration-rotational energy of the final state,
$\Theta_{jK}(\theta)$ is a normalized associated Legendre
polynomial \cite{Zare}, $\Pi_{n(j\,)}^{(in)}(v)$ is the
vibrational wavefunction of the initial diatomic, which satisfies
the following equation:
\begin{eqnarray}
\biggl[-\frac{\,\hbar^2}{2\mu}\frac{d^2}{dv^2} + U^{(in)}(v) +
\frac{\,\hbar^2 j (j+1)}{2 \mu v^2} \biggr]
\Pi_{n(j\,)}^{(in)}(v)=\epsilon_{n(j\,)}^-\,
\Pi_{n(j\,)}^{(in)}(v), \label{23}
\end{eqnarray}
where $U^{(in)}(v)$ is bounding potential of diatom. Similar
expression is valid for $\Pi_{n(j)}^{(out)}(v)$ after replacing
$U^{(in)}(v)\to U^{(out)}(v)$ and
$\epsilon_{n(j\,)}^-\to\,\epsilon_{n(j)}^+ $.

The matching conditions between $\alpha$ and $\beta$ full internal
wavefunctions  can be write in the following form (see for example
\cite{Davydov,MBaer}):
\begin{eqnarray}
\label{24}
 \sum_{\bar{K}}D^J_{M\bar{K}}(\zeta_{1\alpha},\zeta_{2\alpha},\zeta_{3\alpha})
 \rho_\alpha^{-1/2}{\Phi}^{(\pm)J}_{nj\bar{K}}(u_\alpha,v_\alpha,\theta_\alpha)=
 \sum_{\bar{K}}D^J_{M\bar{K}}(\zeta_{1\beta},\zeta_{2\beta},\zeta_{3\beta})\rho_\beta^{-1/2}
{\Phi}^{(\pm)J}_{nj\bar{K}}(u_\beta,v_\beta,\theta_\beta),
\nonumber\\
\rho_{\varsigma}(u_\varsigma,v_\varsigma)=\eta_{\varsigma}
q^2_{0\varsigma}q^2_{1\varsigma}, \qquad
\varsigma=\{\alpha\,\beta\},\qquad \qquad\qquad\qquad\qquad\qquad
\end{eqnarray}
 The relation between the rotation functions in different
body-fixed frames in (\ref{24}) as well know may be represent by
formula \cite{MBaer,Davydov,Rose}:
\begin{equation}
\label{25}
 D^J_{M\bar{K}}(\zeta_{1\beta},\zeta_{2\beta},\zeta_{3\beta})=
 \sum_{K'}d^J_{MK'}(\vartheta)
  D^J_{\bar{K}K'}(\zeta_{1\alpha},\zeta_{2\alpha},\zeta_{3\alpha}).
\end{equation}
 Note that  $\vartheta$ is the angle between vectors
${\bf{q}}_{0\alpha}$ and ${\bf{q}}_{0\beta}$, and may be described
as a function of the internal coordinates using the relation:
\begin{equation}
\label{26}
{\bf{q}}_{0\alpha}{\bf{q}}_{0\beta}={q}_{0\alpha}{q}_{0\beta}
\cos\vartheta,
\end{equation}
where $\vartheta$-angle between $\alpha$ and $\beta$ body-fixed
reaction coordinates systems.

Using the expressions (\ref{24}) and (\ref{25}) it is easy to find
the following  equation of connection (Appendix D):
\begin{eqnarray}
\label{27} \Phi^{(\pm)J}_{njK} (u_\alpha,v_\alpha,\theta_\alpha) =
\sqrt{\frac{\rho_\alpha}{\rho_\beta}}
 \sum_{\bar{K}} d^J_{K\bar{K}}(\vartheta)\Phi^{(\pm)J}_{nj
 \bar{K}}(u_\beta,v_\beta,\theta_\beta)
\end{eqnarray}

Now taking into account that the  overall wavefunction
$\Phi^{(+)J}_{nj\bar{K}} (u_\beta,v_\beta,\theta_\beta)$ at the
limit $u\rightarrow+\infty$ going to clear $(out)$ asymptotic
state (\ref{21}) from equation (\ref{19}) we can fined the
following expression for scattering matrix (see Appendix D):
\begin{eqnarray}
\label{28} \emph{S}^{J}_{n'j\,'K'\,\leftarrow\,njK}(E)=
 \lim_{u\rightarrow+\infty}\Bigl\langle{\Phi}^{(+)J}_{njK}
(u_\alpha,v_\alpha,\theta_\alpha)
{\Phi}_{n'j\,'K'}^{(out)J}(u_\beta,v_\beta,\theta_\beta)
\Bigr\rangle_{v\theta},\nonumber\\
\bigl\langle...\bigr\rangle_{v\theta}=
\int_v\,\eta^{1/2}\,dv\int_{\theta}\sin\theta\,d\theta.
\end{eqnarray}
Putting the expression for $\Phi^{(+)J}_{njK}
(u_\alpha,v_\alpha,\theta_\alpha)$  from (\ref{27}) in the
(\ref{28}) we can find:
\begin{eqnarray}
\label{29} \emph{ S}_{n'j\,'K'\,\leftarrow\,njK}(E)
=\sqrt{\frac{p^{+}_{n'j\,'}}{p^-_{n
j}}}\lim_{u\rightarrow+\infty}\Bigl\langle{\Phi}^{(+)J}_{njK}
(u,v,\theta) {\Phi}_{n'j\,'K'}^{(out)J}(u,v,\theta)
\Bigr\rangle_{v\theta}.
\end{eqnarray}
 In the expression (\ref{29})
 $ \emph{S}_{n'j\,'K'\,\leftarrow\,njK}$
  maybe interpreted as a general form for the reactive
$\emph{\textbf{S}}$-matrix elements.
 Evidently from definition of the  $\emph{\textbf{S}} $-matrix
elements (\ref{29}) the integration over coordinate $u$ is absent,
however it is easy to understand that in the $(out)$ asymptotic
region the full phase is cancelled out  and thus
$\emph{S}_{n'j\,'K'\,\leftarrow\,njK}$ -matrix elements become
independent from coordinate $u$.

 Now the differential reactive state-to-state cross section
may be simply constructed  with the help of matrix elements
$\emph{S}^{J}_{n'j\,'K\,\leftarrow\,njK}(E)$ (see particularly
works \cite{Schatz76,Kurti,GBK}):
\begin{eqnarray}
\sigma_{n'j\,'\,\leftarrow\,nj}(E,\vartheta)=
\frac{1}{2j+1}\sum_{KK'}\frac{1}{4p^2_{nj}}\biggl|\sum_J(2J+1)
\emph{S}_{n'j\,'K'\,\leftarrow\,njK}(E)\,
d^J_{KK'}(\vartheta)\biggl|^2,
 \label{30}
\end{eqnarray}
where $p_{nj}$ is the incoming wave vector.

The total integral reactive  cross section for reaction from a
particular initial state to  all possible final states is then
given by a summation over all total angular momenta which can
contribute to the reaction:
\begin{equation}
\sigma^{tot}_{all\leftarrow\,nj}=\frac{\pi}{p^2_{nj}}\sum_{J}\,(2J+1)
P^J_{r}(E),
 \label{31}
\end{equation}
where
\begin{equation}
\qquad P^J_{r}(E)=\frac{1}{2j+1}\sum_{n'j\,'}\sum_{KK'}
\Bigl|\emph{S}_{n'j\,'K'\,\leftarrow\,njK}(E)\Bigl|^2.
 \label{32}
\end{equation}
Recall that  $P^J_{r}(E)$  total reaction probability for a
particular value of the total angular  momentum $J$.

So, we have obtained a new representation (\ref{29}) for the
scattering $\emph{\textbf{S}}$-matrix, where the coordinate $u$,
which varies along the reaction curve $\Im_{if}$, plays a role
much as usually time does in standard quantum scattering theory.
The reaction coordinate $u$ could thus be viewed as an
\emph{intrinsic time} of the scattering.

\subsection{Coupled-channel expression for full wavefunction and
 $\emph{\textbf{S}}$-matrix elements }

Remembering that along $u$ the system is in translational motion,
while along $v$ and $\theta$ the motion is localized we employ the
time-independent coupled-channel (CC) approach.

It is convenient to write down the intrinsic full wavefunction  in
the following  form (see for example \cite{Light2}):
\begin{equation}
{\Phi}^{(+)J}_{K'[\varrho]}(u,v,\theta)=\sum_{\bar{n}\bar{j}}
{G}^{(+)J}_{\bar{n}\bar{j}K'\,[\varrho]}(u)
\Xi_{\bar{n}(\bar{j})}(v;u)\Theta_{\bar{j}K'}(\theta),\quad
\varrho=(njK),
 \label{33}
\end{equation}
and necessitate the wavefunction (\ref{33}) to satisfy to initial
conditions (\ref{20}), the symbol $[\varrho]$ describes a set of
initial quantum numbers. Note that it is standard way to obtain
the close system of coupled equations for scattering (or
translational) functions ${G}^{(+)J}_{\varrho\,'\,[\varrho]}(u)$.

 The vibrational part of the wavefunction $\Xi_{n(j)}(u,v)$
forms an orthonormal basis in the variable $v$ for each fixed
value of $u$ and satisfies the equation:
\begin{eqnarray}
\biggl[-\frac{\,\hbar^2}{2\mu}\frac{d^{\,2}}{dv^2} +
\overline{U}(u,v) + \frac{\hbar^2j(j+1)}{2\mu{v^2}}
 \biggr]
\Xi_{n(j)}(v;u)=\epsilon_{n(j)}(u)\,\Xi_{n( j)}(v;u), \label{34}
\end{eqnarray}
where
\begin{eqnarray}
\label{35} \overline{U}(u,v) =
U(u,v,\theta)\bigl|_{\theta=0}-U_{eff}(u,v).
\end{eqnarray}
Recall, that $U(u,v,\theta)\bigl|_{\theta=0}$ corresponds to the
potential energy of collinear collision.

 In some situations it is useful to approximate
$\overline{U}(u,v)$ such that Eq. (\ref{34}) can be analytically
solved (see section E). Note that the (\ref{34}) in the limit
$u\to-\infty$ transforms exactly into the asymptotic equation
(\ref{23}). Correspondingly in the limit of $u\to+\infty$ Eq.
(\ref{34}) describing bound state of the $(out)$ asymptotic
channel. Note that for both cases $u\to \mp\infty$ the effective
potential energy $U_{eff}(u,v)$ limits to zero.

For subsequent analytical manipulations we give two important
expressions for Legendre polynomials \cite{Edmonds,Zare}:
\begin{equation}
\biggl[\frac{1}{\sin\theta}\frac{\partial}{\partial{\theta}}\sin\theta
\frac{\partial}{\partial{\theta}}
-\frac{K^2}{\sin^2\theta}\biggr]\Theta_{jK}(\theta)=-
j(j+1)\Theta_{jK}(\theta),
 \label{36}
\end{equation}
and
\begin{equation}
\biggl[\mp\frac{\partial}{\partial{\theta}}+ (K\pm
1)\cot\theta\biggr] \Theta_{j(K\pm
1)}(\theta)={c^{\pm}_{jK}}\Theta_{jK}(\theta),\quad
c^{\pm}_{jK}=\bigl[j(j+1)-K(K\pm1)\bigr]^{1/2}.
 \label{37}
\end{equation}
Next substitute the wavefunction expression (\ref{33}) into Eq.
(\ref{17}) taking into account (\ref{34})-(\ref{37}) and multiply
by $\Xi_{n'(j\,')}(v;u)$ and $\Theta_{j\,'K'}(\theta)$. Thereafter
integrate over the angle $\theta$ and coordinate $v$ to find the
following equation:
\begin{eqnarray}
\Biggl\{\biggl[\delta_{n'\bar{n}}\frac{d^2}{d{u^2}}+2\biggl<\frac{\partial}{\partial{u}}
-\frac{1}{\eta}\frac{\partial\eta}{\partial{u}}\biggr>_{n'\bar{n}}\frac{d}{d{u}}
+\biggl<\frac{\partial^2}{\partial{u^2}} -\frac{2}{\eta}
\frac{\partial\eta}{\partial{u}}\frac{\partial}{\partial{u}}\biggr>_{n'\bar{n}}
+\frac{2\mu}{\hbar^2}\biggl<\eta^2\Bigl[E-E_{JK'}(u,v)
\nonumber\\
-\epsilon_{\bar{n}(\bar{j\,})}(u) +U(u,v)
+\frac{\hbar^2\bar{j\,}(\bar{j\,}+1)}{2\mu{v^2}}\Bigr]\biggl>_{n'\bar{n}}\biggr]
\delta_{j\,'\bar{j\,}}-\frac{2\mu}{\hbar^2}\Bigl<\eta^2
U_{j\,'\bar{j\,}}^{K'}(u,v)\Bigr>_{n'\bar{n}}\Biggr\}{G}^{(+)J}_{\bar{n}\bar{j\,}K'\,[\varrho]}(u)
\nonumber\\
+\Bigl<\frac{\eta^2}{q_0^2}\Bigr>_{n'\bar{n}}\delta_{j\,'\bar{j\,}}
\left[C_{J\bar{j\,}\,K'}^+{G}^{(+)J}_{\bar{n}\bar{j\,}\,(K'+1)\,[\varrho]}(u)
+C_{J\bar{j\,}\,K'}^-{G}^{(+)J}_{\bar{n}\bar{j\,}\,(K'-1)\,[\varrho]}(u)\right]
 = 0,\qquad
 \label{38}
\end{eqnarray}
where $C^{\pm}_{JjK}=c^{\pm}_{JK}c^{\pm}_{jK}$ (see expressions
(\ref{11}) and (\ref{37})\,).

The equation (\ref{38}) may be presented in another form:
\begin{eqnarray}
\Biggl\{\biggl[\delta_{n'\bar{n}}
\frac{d^2}{d{u^2}}+2\biggl<\frac{\partial}{\partial{u}}
-\frac{1}{\eta}\frac{\partial\eta}{\partial{u}}\biggr>_{n'\bar{n}}\frac{d}{d{u}}
+\biggl<\frac{\partial^2}{\partial{u^2}} -\frac{2}{\eta}
\frac{\partial\eta}{\partial{u}}\frac{\partial}{\partial{u}}\biggr>_{n'\bar{n}}
+\frac{2\mu}{\hbar^2}\biggl<\eta^2\Bigl[E-E_{J\bar{K}}(u,v)\,\,
\nonumber\\
-\epsilon_{\bar{n}(\bar{j}\,)}(u)+ U(u,v)+
\frac{\hbar^2\bar{j\,}(\bar{j\,}+1)}
{2\mu{v^2}}\Bigr]\,\biggl>_{n'\bar{n}}\,\,\biggr]\delta_{j\,'\bar{j\,}}\delta_{K'\bar{K}}
 -\frac{2\mu}{\hbar^2}\Bigl<\eta^2\,U_{j\,'\bar{j\,}}^{\bar{K}}(u,v)\Bigr>_{n'\bar{n}}\,
\delta_{K'\bar{K}}\quad
\nonumber\\
+\Bigl<\,\frac{\eta^2}{q_0^2}\,\Bigr>_{n'\bar{n}}\,\delta_{j\,'\bar{j\,}}
\left[\delta_{K'+1\,\bar{K}}\,{C_{J\bar{j\,}(\bar{K}-1)}^+
+\delta_{K'-1\,\bar{K}}\,C_{J\bar{j\,}(\bar{K}+1)}^-}\right]
\Biggr\}{G}^{(+)J}_{\bar{n}\bar{j\,}\bar{K}\,[\varrho]}(u)
=0,\qquad\quad
 \label{39}
\end{eqnarray}
 where the summation over repeating index $n'$ and $j\,'$
 are implied and we
 use the following notation for matrix elements:
\begin{eqnarray}
\bigl<f(u)\bigr>_{nn'}=\int^{+\infty}_{-\infty}
\Xi_{n(j)}(v;u)f(u,v)\Xi_{n'(j)}^\ast(v;u)dv,
\nonumber\\
U^K_{jj\,'}(u,v)=\int^{\pi}_0{\Theta_{jK}(\theta)\,U(u,v,\theta)
\, \Theta_{j' K}(\theta)}\sin\theta{d\theta}.
 \label{40}
\end{eqnarray}

Thus we obtain a system of $N$ coupled second-order differential
equations (\ref{38}) or (\ref{39}), which can be rewritten in a
form of $2N$ coupled first-order ordinary differential equations
system:
\begin{eqnarray}
\frac{d}{du}\textbf{F}=\biggl(\begin{matrix}
0 \,\,\,\,{\bf{I}}\\
{\bf{A}}\,\, {\bf{B}} \\
\end{matrix}\biggr)
\textbf{F},
 \label{41}
\end{eqnarray}
where
\begin{eqnarray}
[\textbf{F}(u)]_{i\leq{N}}={G}^{(+)J}_{\bar{n}\bar{j\,}\bar{K}
\,[\varrho]}(u),\quad
[\textbf{F}(u)]_{i>{N}}=\frac{d{G}^{(+)J}_{\bar{n}\bar{j\,}\bar{K}
\,[\varrho]}(u)}{du},\quad {\bf{I}}=\left(\begin{matrix}
1 \,...\,0\\
.\,1\,.\\
0\,...\,1 \\
\end{matrix}\right).
 \label{42}
\end{eqnarray}

Moreover in Eq. (\ref{41}) the following denotations are made:
\begin{eqnarray}
A_{n'j\,'K'\,|\,\bar{n}\bar{j\,}\bar{K}}(u)=\biggl[
\biggl<\frac{\partial^2}{\partial{u^2}} -\frac{2}{\eta}
\frac{\partial\eta}{\partial{u}}\frac{\partial}{\partial{u}}\biggr>_{n'\bar{n}}
+\frac{2\mu}{\hbar^2}\biggl<\eta^2\Bigl[E-E_{J\bar{K}}(u,v)-
\epsilon_{\bar{n}(\bar{j\,})}(u) +U(u,v)
\nonumber\\
 +\,\frac{\hbar^2\bar{j\,}(\bar{j\,}+1)}
{2\mu{v^2}}\Bigr]\biggl>_{n'\bar{n}}\,
-\,\Bigl<\eta^2\,U_{j\,'\bar{j\,}}^{\bar{K}}(u,v)\Bigr>_{n'\bar{n}}
\biggr]\delta_{j\,'\bar{j\,}}\delta_{K'\bar{K}}
\nonumber\\
+\Bigl<\,\frac{\eta^2}{q^2_0}\,\Bigr>_{nn'}\,\delta_{j\,'\bar{j\,}}
\left[\delta_{K'+1\,\bar{K}}\,{C_{J\bar{j\,}(\bar{K}-1)}^+}
+\delta_{K'-1\,\bar{K}}\,{C_{J\bar{j\,}(\bar{K}+1)}^-}\right],
\nonumber\\
B_{n'j\,'K'\,|\,\bar{n}\bar{j\,}\bar{K}}(u)=-2\biggl<\frac{\partial}{\partial{u}}
-\frac{1}{\eta}\frac{\partial\eta}{\partial{u}}\biggr>_{n'\bar{n}}\,
\delta_{j\,'\bar{j\,}}\,\delta_{K'\bar{K}},
\qquad\qquad\qquad\qquad\qquad\qquad\qquad\quad\qquad
 \label{43}
\end{eqnarray}

Now we turn to the derivation of $\emph{\textbf{S}}$-matrix
elements using the representation  for the full wavefunction
(\ref{33}).

As it is  known the full wavefunction of three-body system is
determined by the set of three quantum numbers $(n,j,K)$. If we
want to describe the full wavefunction of body system with the
help of representation (\ref{33}) it is necessary to demand the
asymptotic condition:
\begin{equation}
\label{44}
\lim_{u\rightarrow-\infty}\sum_{n'j\,'}{G}^{(+)J}_{n'j\,'K'\,
[\varrho]}(u) = \frac{1}{\sqrt{2 \pi}} \exp \left(-i p_{n j}^-\,u
\right) \delta_{n n'} \delta_{j j\,'}\delta_{K K'},
\end{equation}
 to be fulfilled. It means
that in the limit  $u\to-\infty$, the wavefunction (\ref{33})
transforms to the asymptotic wavefunction
${\Phi}_{njK}^{(in)J}(u,v,\theta)$ (see formula (\ref{20})\,).

The $\emph{\textbf{S}}$-matrix elements  are obtained after
substituting (\ref{33}) into the expression (\ref{30}):
\begin{eqnarray}
\emph{S}_{n'j\,'K'\,\leftarrow\,njK}(E)=
\sqrt{\frac{p^{+}_{n'j\,'}}{p^-_{nj}}}
\lim_{u\rightarrow+\infty}\sum_{\bar{n}\bar{j}}
 G^{(+)J}_{\bar{n}\bar{j}K'\,[\varrho]}(u)\,W_{\bar{n}n'}(u)
 \Lambda_{\bar{j}K'\,\leftarrow\,jK'},
 \label{45}
 \end{eqnarray}
 where
\begin{eqnarray}
W_{n'\bar{n}}(u)=\Bigl\langle\Xi_{\bar{n}(\bar{j\,})}(v;u)
\Pi_{n'(j\,')}^{(out)}(v)\Bigr \rangle_{v}, \qquad
\Lambda_{\bar{j}K'\,\leftarrow\,jK'}=
\Bigl<\Theta_{\bar{j\,}K'}(\theta)\Theta_{jK'}
(\theta)\Bigr>_{\theta}=\delta_{\bar{j\,}j}.
 \label{46}
\end{eqnarray}
The expression for $\textbf{\emph{S}}$-matrix elements (\ref{45})
can be simplified, if we take as basis the functions
$\Xi_{n(j)}(v;u)$, which in the limit $u\rightarrow+\infty$
coincide with the orthonormal basic wavefunctions
$\Pi_{\bar{n}(j)}^{(out)}(v)$. In this case we get the
simplification
$\lim_{u\rightarrow+\infty}W_{n'\bar{n}}(u)=\delta_{n'\bar{n}}$
and the following expression holds for $\textbf{\emph{S}}$-matrix
elements:
\begin{eqnarray}
\emph{S}_{n'j\,'K'\,\leftarrow\,njK}(E)
=\sqrt{\frac{p^{+}_{n'j\,'\,}} {p^-_{nj}}}\,
 {G}^{(+)J}_{\varrho'\,[\varrho]}(E;+\infty),\qquad \varrho=(njK).
 \label{47}
\end{eqnarray}

 So, we have now shown that the initial quantum
scattering problem can be exactly reduced to the standard set of
$N$ coupled second order differential equations
(\ref{38})-(\ref{39})  in a single variable $u$ (where $N$ is the
number of channels or coupled equations). For the solution of this
system of equation it is useful to represent it to the form
(\ref{41}). The matrix equation (\ref{41}) has 2$N$ linearly
independent solutions. There are several standard ways in which
this set of equations may be solved
\cite{Lester,Manolopoulos,MillerAnnRev,LightWalker}. The $2N$
independent solutions are then combined to give $N$ solutions
which obey the asymptotic conditions as set out in Eqs.
(\ref{44}). The process of combining the independent solutions to
satisfy the boundary conditions automatically yields the
$\emph{\textbf{S}}$-matrix elements.

\subsection{Etalon equation method, full wavefunction and
 $\emph{\textbf{S}}$-matrix elements }

During the development of the algorithm for numerical simulation
based on aforementioned theoretical approach it is necessary to
recover the scattering matrix equation (\ref{41})-(\ref{43}) on
the grid. The grid is formed along the reaction coordinate curve,
which connects $(in)$ and $(out)$ scattering channels. The problem
is that at each point of the grid one has to solve $M_i=n_i\times
j_i$ number of one-dimensional quantum problem and by further
numerical integration to find the form of the matrix equation
(\ref{41}) on the grid. This procedure requires huge amount of
computer resource and undergoes the accumulation of computation
errors. Besides matrix equation (\ref{41}) in this case will be
given as a numerical array, which is extremely inconvenient for
numerical simulation with changing integration scale. In order to
overcome this difficulty one can use the exactly solvable model
for vibration states in the total wavefunction coupled-channel
expression. This allows getting analytical form for
one-dimensional matrix equation of reactive scattering.

So, for full wavefunction can be written another coupled-channel
representation:
\begin{equation}
{\Phi}^{(+)J}_{K'\,[\varrho]}(u,v,\theta)=\sum_{\bar{\nu\,}\bar{j\,}}
{G}^{(so)J}_{\bar{\nu\,} \bar{j\,}K'\,[\varrho]}(u)
\widetilde{\Xi}^{(so)}_{\bar{\nu\,}(\bar{j\,})}(u,v)\Theta_{\bar{j\,}K'}(\theta),
\qquad 0\leq\bar{\nu\,}<\infty,
 \label{48}
\end{equation}
where $\widetilde{\Xi}^{(so)}_{\nu (j)}(u,v)$ is the wavefunction
of singular nonstationary (on reaction coordinate $u$, or later
\emph{intrinsic time} $\tau$) quantum oscillator and satisfying
the following equation \cite{MalMan} (later will be named the
etalon equation):
\begin{eqnarray}
i\hbar\frac{\partial{\Xi^{(so)}}}{\partial{\tau}}=
\biggl[-\frac{\,\hbar^2}{2\mu}\frac{d^2}{dv^2} +
\biggl(\frac{\mu\,\omega^2(\tau)}{2}v^2 + \frac{g}{{v^2}}\biggr)
 \biggr]
\Xi^{(so)}, \qquad \Xi^{(so)}(\tau;v)\equiv \widetilde{\Xi}^{(so)}(u,v),\nonumber\\
\tau=u/\upsilon_0,\qquad
 g=\frac{\hbar^2}{2\mu}j(j+1),\qquad
0<{v}<\infty, \quad -\infty<u,\tau<+\infty. \qquad\label{49}
\end{eqnarray}
Note that $\upsilon_{0}=a_0\omega_{i}$ (where $a_0$ is a Bhor
radius, for explanation $\omega_i$ see (\ref{53})) give the
characteristic speed by coordinate of localized motion $v$.

The Eq. ({\ref{49}}) can be solved exactly \cite{Gamiz} for values
$v\ge0$:
\begin{eqnarray}
\label{50}
\Xi^{(so)}_{\nu(j)}(\tau;v)=\biggl[2\biggl(\frac{\mu}{\hbar{\sigma^2}}
\biggr)^{a+1}\frac{\Gamma(n+1)}{\Gamma(n+a+1)}\biggr]^{1/2}
 v^{a+1/2}\exp\Bigl(-2i\nu\gamma
+i\frac{\mu}{2\hbar}\frac{\dot{\sigma}}{\sigma}v^2\Bigr)
L_{\nu}^a\Bigl(\frac{\mu\dot{\gamma}}{\hbar}v^2\Bigr),\nonumber\\
\dot{\sigma}=\frac{d\sigma}{d\tau},\qquad
\dot{\gamma}=\frac{d\gamma}{d\tau},\qquad
a=\frac{1}{2}\Bigl(1+\frac{8\mu{g}}{\hbar^2}\Bigr)^{1/2}.
\qquad\qquad\qquad\qquad
\end{eqnarray}
It is significant that the solution (\ref{50}) may be analytically
continued to the infinite numerical axis $-\infty<v<+\infty$ (see
particularly (\cite{MalMan})). In  (\ref{50}) the function
$\gamma(\tau)=\int^u{|\sigma(\tau)|^{-2}}{d\tau}$ is the argument
of complex function $\sigma$ which is a solution of classical
oscillator problem:
\begin{equation}
\label{51} \ddot{\sigma}+\omega^2(\tau)\sigma=0.
\end{equation}
For further investigations it is useful to have an analytically
solvable model for function $\sigma(\tau)$. In particularly  the
 Eq. (\ref{51}) may be integrated exactly if the frequency
 is described by following model form (see Appendix E) \cite{Witten}:
\begin{equation}
\label{52} \omega^2(\tau)=A_0+A_1\tanh(\lambda{\tau}),
\end{equation}
where $A_0$, $A_1$ and $\lambda$ are some adjusting constants,
which can be chosen to be most suitable for numerical simulations.
 As it is evident from formula (\ref{52}),
$\lim_{u,\tau\rightarrow\mp\infty}\omega^2(\tau)=\omega^2_{i,f}$
and corresponding asymptotic solutions of Eq. (\ref{51}) are:
\begin{eqnarray}
\label{53} \sigma(\tau)\,\,\stackrel{\sim}{ _{\tau \rightarrow
-\infty}}\,\, \omega^{-1/2}_{i}e^{i\omega_{i}\tau},
\qquad\qquad\qquad\qquad\qquad\qquad\qquad\qquad
\nonumber\\
\sigma(\tau)\,\,\stackrel{\sim}{ _{\tau \rightarrow +\infty}}\,\,
\omega^{-1/2}_{f}\Bigl(C_{1}e^{i\omega_{f}\tau}-C_{2}
e^{-i\omega_{f}\tau}\Bigr),\qquad\quad\quad
\omega^2_{i,f}=A_0\mp{A_1},
\end{eqnarray}
where $\omega_{i,f}$ are coefficients at second terms of expansion
series in corresponding asymptotic bound states potential energies
(see Eq. (\ref{23})). The coefficients $C_{1(2)}$ are some complex
numbers which are found  by the solution of Eq. (\ref{51}) in the
limit $\tau\to+\infty$. The constraint $|C_1|^2-|C_2|^2=1$ is
hashed at the numbers $C_1$ and $C_2$, which is followed by
commutation correspondence.

Using this fact it is easy to calculate bound-state energy in
asymptotic channels
$\epsilon_{\nu(j)}=\hbar\omega_{i,f}(2\nu+a+1)$ and the asymptotic
form of the etalon wavefunction (\ref{50}). In particular, in the
$(in)$
 asymptotic channel:
\begin{eqnarray}
\label{54}\Xi^{(so)}_{\nu(j)}(\tau;v)\,\,\stackrel{\sim}{ _{\tau
\rightarrow -\infty}}\,\,\Xi^{(-)}_{\nu(j)}(\tau;v)=
\qquad\qquad\qquad\qquad\qquad\qquad\qquad\qquad\qquad\qquad\nonumber\\
\biggl[2\biggl(\frac{\mu\,\omega_{i}}{\hbar}\biggr)^{a+1}
\frac{\Gamma(n+1)}{\Gamma(n+a+1)}\biggr]^{1/2}
 v^{a+1/2}\exp\Bigl(-2i\nu{\omega_{i}\tau}
-\frac{\mu\omega_{i}}{2\hbar}v^2\Bigr)
L_{\nu}^a\Bigl(\frac{\mu\omega_{i}}{\hbar}v^2\Bigr).
\end{eqnarray}

 In the
Eq. (\ref{54}) $L_{\nu}^a({\hbar}^{-1}{\mu\dot{\gamma}}v^2)$ is
Lager polynomial and  the wavefunctions
$\Xi^{(so)}_{\nu(j)}(\tau;v)$ form
 an orthonormal basis:
$\int_0^{\infty}\Xi^{(so)}_{\nu(j)}(\tau;v)\Xi^{(so)\ast}_{\lambda(j)}(\tau;v)
dv= \delta_{\nu\lambda}$.

 When  $g=0$ $(a=1/2)$ the wavefunctions (\ref{54})  describe
 the nonstationary oscillator with wall at the beginning of coordinate
 \cite{DoMaMal}. It is well know that the wavefunction of this system
 coincides
 with odd functions of harmonic oscillator in view of identity
 $H_{2\nu+1}(x)=(-1)^{\nu}2^{2\nu+1}{\nu}!xL_{\nu}^{1/2}(x^2)$ (see for example \cite{Erd}).
 Nevertheless the Eq. (\ref{49}) have another type solution too,
 which coincides with even wavefunctions of harmonic oscillator \cite{Leb}.

 So, in case of $g=0$ the solution of etalon equation (\ref{49})
 is  wavefunction of quantum harmonic oscillator
 depending on time (in this case  \emph{intrinsic
 time} $\tau$)  frequency \cite{Husimi,MalMan}:
\begin{eqnarray}
\label{55}\Xi^{(so)}_{\nu}(\tau;v)=\biggl(\frac{\dot{\sigma}^{\ast}}
{2\sigma}\biggr)^{\nu/2}({\nu}!\sigma\sqrt{\pi})^{-1/2}
\exp\Bigl(i\frac{\dot{\sigma}}{2\sigma}v^2\Bigr)
H_{\nu}\Bigl(\frac{v}{|\sigma|}\Bigr).
\end{eqnarray}

Now comparing representations (\ref{34}) and (\ref{48}) is easy to
find the following relation between reactive scattering functions
${G}^{(+)J}_{\bar{n}\bar{j\,}K'\,[\varrho]}(u)$ and
${G}^{(so)J}_{\bar{n}\bar{j}K'\,[\varrho]}(u)$:
\begin{equation}
\label{56}
{G}^{(+)J}_{\bar{n}\bar{j\,}K'\,[\varrho]}(u)=\sum_{\bar{\nu}}\,
C_{\bar{\nu}{\bar{n}}(\bar{j\,})}(u){G}^{(so)J}_{\bar{\nu}
\bar{j\,}K'\,[\varrho]}(u),
\end{equation}
where matching coefficients $C_{\bar{\nu}{\bar{n}}(\bar{j\,})}(u)$
is defined by formula:
\begin{equation}
\label{57}
C_{\bar{\nu}{\bar{n}}(\bar{j\,})}(u)=\int_{-\infty}^{+\infty}\Xi^{(so)}_{\bar{\nu}(\bar{j\,})}
(u\upsilon_0^{-1};v)\Xi^{\ast}_{\bar{n}(\bar{j\,})}(u,v)dv.
\end{equation}
Using (\ref{57}) in the limit $u\rightarrow-\infty$ for definition
of asymptotic conditions for the  function
${G}^{(so)J}_{\bar{\nu}\bar{j\,}K'\,[\varrho]}(u)$ the system of
linear equations can be obtained:
\begin{eqnarray}
  \begin{cases}
\sum_{\bar{\nu}}\,C^{-}_{\bar{\nu}{0}(\bar{j\,})}(u)\,
{G}^{(so)J}_{\bar{\nu}{\bar{j\,}K'\,[\varrho]}}(u)=0,\\
\quad\qquad\qquad...\quad\qquad\qquad\,,\\
\sum_{\bar{\nu}}\,C^{-}_{\bar{\nu}{\bar{n}}(\bar{j\,})}(u)\,
{G}^{(so)J}_{\bar{\nu}{\bar{j\,}K'\,[\varrho]}}(u)=o(u),\\
\quad\qquad\qquad...\quad\qquad\qquad\,,\\
\sum_{\bar{\nu}}\,C^{-}_{\bar{\nu}{\bar{m}}(\bar{j\,})}(u)\,
{G}^{(so)J}_{\bar{\nu}{\bar{j\,}K'\,[\varrho]}}(u)=0,
 \end{cases}
\label{58}
\end{eqnarray}
 where $m$  defines the number of vibrational states in the
$(in)$ channel in the same way as
$o(u)={\bigl(2\pi\bigr)}^{-1/2}\exp\bigl(-ip^{-}_{nj}u\bigr)$ and
$C^-_{\bar{\nu}{\bar{n}}(\bar{j\,})}(u)=
\int_{-\infty}^{+\infty}\Xi^{(-)}_{\bar{\nu}(\bar{j\,})}
(u\upsilon_0^{-1};v)
\widetilde{\Xi}^{\ast}_{\bar{n}(\bar{j\,})}(u,v)dv$.

In the  system  (\ref{58}) the number of equations is finite while
the number of unknown quantities is infinite. For exact definition
of problem (\ref{58}) we have to demand following conditions:
 \begin{eqnarray}
\label{59}
{G}^{(so)J}_{\bar{\nu}{\bar{j\,}K'\,[\varrho]}}(u)\equiv 0, \quad
\nu>m.
\end{eqnarray}

Now substituting  (\ref{48}) into  (\ref{17}) and after simple
analytical calculation we can get a new equation for reactive
scattering:
\begin{eqnarray}
&&\Biggl\{\delta_{j\,'\bar{j\,}}\,\delta_{K'\bar{K}}\biggl[\delta_{\nu\,'\bar{\nu}}
\frac{d^2}{d{u^2}}+2\biggl<\frac{\partial}{\partial{u}}
-\frac{1}{\eta}\frac{\partial\eta}{\partial{u}}\biggr>_{\nu\,'\bar{\nu}}
\frac{d}{d{u}}+\biggl<\frac{\partial^2}{\partial{u^2}}
-2\biggl(\frac{1}{\eta}\frac{\partial\eta}{\partial{u}}
+i\frac{\mu}{\hbar}\upsilon_{0}\biggr)
\frac{\partial}{\partial{u}}\biggr>_{\nu\,'\bar{\nu}}
\nonumber\\
&&\quad+\frac{2\mu}{\hbar^2}\biggl<\eta^2\Bigl[E-E_{J\bar{K}}(u,v)
-U_{eff}(u,v)+\frac{1}{2}\mu\omega^2v^2+\frac{\hbar^2\bar{j\,}(\bar{j\,}+1)}
{2\mu{v^2}}\Bigr]\biggl>_{\nu\,'\bar{\nu}}\,\biggr] \nonumber
\\
&&\qquad+\delta_{\bar{j\,}j\,'}
\biggl<\frac{\eta^2}{q^2_0}\biggr>_{\nu\,'\bar{\nu}}
\Bigl[\delta_{K'+1\,\bar{K}}\,C_{J\bar{j\,}\bar{K}-1}^+
+\delta_{K'-1\,\bar{K}}\,C_{J\bar{j\,}\bar{K}+1}^-\Bigr]
\nonumber\\
&&\qquad-\frac{2\mu}{\hbar^2}\Bigl<\eta^2
U_{j\,'\bar{j\,}}^{\bar{K}}(u,v)\Bigr>_{\nu\,'\bar{\nu}}\delta_{K'\bar{K}}\Biggr\}
{G}^{(so)J}_{\bar{\nu} \bar{j\,}\bar{K}\,[\varrho]}(u) =0.
 \label{60}
\end{eqnarray}
 In this case the summation over repeating index $\bar{\nu}$
is implied and we use the following notation for matrix elements:
\begin{eqnarray}
\bigl<f(u)\bigr>_{\nu\nu\,'}=\int^{+\infty}_{-\infty}\Xi^{so}_{\nu(j)}
(u\upsilon_0^{-1};v)f(u,v)\Xi^{{so}^\ast}_{\nu\,'(j)}(u\upsilon_0^{-1};v)dv.
 \label{61}
\end{eqnarray}
Remind that the system of second order differential Eq. (\ref{60})
may be written in the first order form of type
(\ref{41})-(\ref{43}).

Finally, taking into account (\ref{44})-(\ref{46}) (\ref{48}) and
(\ref{57})  one can find the new expression for
$\textbf{\emph{S}}$-matrix elements:
 \begin{eqnarray}
\label{62} \emph{S}_{n'j\,'K'\,\leftarrow\,njK}(E)=
\sqrt{\frac{p^{+}_{n'j\,'}} {p^-_{nj}}}\,
\lim_{u\rightarrow+\infty}\sum_{\bar{\nu}\,
\bar{n}\,\bar{j\,}}G^{(so)}_{\bar{\nu}\bar{j\,}K'\,[\varrho]}(u)
 C_{\bar{\nu}\bar{n}(\bar{j\,})}(u)W_{\bar{n}n'}(u)
\Lambda_{j\,'K'\,\to\bar{j}\,K'}\nonumber\\
=\sqrt{\frac{p^{+}_{n'j\,'}} {p^-_{nj}}}\,\sum_{\bar{\nu}}
G^{(so)}_{\bar{\nu}{j\,'}K'\,[\varrho]}(E;+\infty)
 C_{\bar{\nu}n'(j\,')}(+\infty).
\end{eqnarray}

So, we found a new analytical expression for the
$\emph{\textbf{S}}$ -matrix elements of reactive scattering
(\ref{62}) with the help of exactly solvable etalon equation
(\ref{49}) and we will call it \emph{etalon equation method}.
Obviously this method is favorable for numerical simulation.
Particularly computation of $\textbf{\emph{S}}$-matrix elements
(\ref{62}) in this case is relatively simple too.

\section{Conclusion}
The introduction of  \emph{natural collision coordinates}  by
Marcus \cite{Marcus} was intended to simplify quantum reactive
scattering calculations. In spite of  significant efforts
\cite{Light,WLAJCP76,LightW} the application of this method to
$3D$ reactive scattering has encountered considerable
difficulties, and as a consequence the investigations in this
direction ceased in 1988 (see for example report
\cite{Manolopoulos}). In particular, the following two problems
can be indicated:
\begin{enumerate}
\item  For an atom-diatom arrangement reaction $A+BC$ in $3D$, at
every conserved total angular momentum $J$, the vibrational
coordinate $v$ (see FIG. 3) becomes a $2D$ (ro-vibrational)
surface in the $3D$ intrinsic space which specifies the size and
the shape of three-atom triangle. It is significant that this
surface has extremely complicated the metric properties and can be
investigated with the help of difficult numerical calculations
\cite{SWL}.
 \item The third arrangement reaction $AC+B$ (see schema 1)
must also be included in the $3D$ computation schema. However,
this leads to problems in the understanding of translational
\emph{reaction coordinate} idea. Moreover, technical difficulties
arise  by \emph{matching surfaces} between the three arrangement
channels.
\end{enumerate}
Nevertheless the mentioned difficulties either can be overcame in
the intrinsic $3D$ space or they do not play an essential role in
the framework of this consideration.

Hence, if we define three different curves of the \emph{reaction
coordinate} $\Im_{if}$ analogously to (\ref{05}) by the way they
connect different asymptotic subspaces ( different reactant and
product channels) one can describe all the scattering channels
(see schema 1). From the other side, by defining this curve and
correspondingly NCC system carefully on the case of collinear
collision  all possible configuration of reactive scattering may
be described.

 In recent articles \cite{Ash,Ash1,Ash2}, one of the authors
together with colleagues have analyzed in detail the difficulties
arising in the formulation of quantum scattering theory in NCC
system.

In particular, it was shown for the case of a collinear scattering
process in NCC system that a procedure similar to that described
above for the $3D$ case could effectively yield the exact
$\emph{\textbf{S}}$-matrix.

In this article the quantum reactive scattering theory in
curvilinear reaction coordinates has been generalized to the $3D$
case.

We have started with Schr\"odinger equation of thee-body system in
Jacobi coordinates (\ref{10}). Using the coordinate transformation
given in Eqs. (\ref{05})-(\ref{08}) this equation has been
transformed into the NCC system. The reaction coordinate $u$ (or
$\tau$) may be considered as a chronological parameter in the
theory (i.e. \emph{intrinsic time}), and plays a role analogous to
that of time in standard time-dependent scattering theory.
Representing the full wavefunction of three-body system by
standard coupled-channel form (\ref{33}) the initial $3D$
multi-channel quantum scattering problem (\ref{17}) with
correspondingly initial and border conditions (\ref{20}) may be
reduced to an inelastic single-arrangement problem Eq.
(\ref{38})-(\ref{39}) or (\ref{41})-(\ref{43}). As a result of
this reducing after the solution of system of Eq.
(\ref{41})-(\ref{43}) on a set of points $u$ of curve of reaction
coordinate $\Im_{if}$ (grid) the full wavefunction and all
transition $\emph{\textbf{S}}$-matrix elements are found
simultaneously (see expression for $\emph{\textbf{S}}$-matrix
elements ({\ref{47}})).

Another direction of investigation of Schr\"odinger Eq. (\ref{17})
is the representation of solution for full wavefunction of
three-body system in the coupled-channel form (\ref{48}) combined
with the exact solvable \emph{etalon equation method} (see Eqs.
(\ref{49})-(\ref{50})). Recall that etalon equation describes the
localization properties of full wavefunction along the curve of
reaction coordinate $\Im_{if}$. This method reduces the $3D$
quantum multichannel scattering problem to inelastic
single-arrangement problem too. Still, the initial conditions for
the system of differential equations (\ref{60}) in this case are a
little bit complicated and can be found by the solution of linear
algebraic equations system (\ref{58})-(\ref{59}).

Note that this is an important theoretical result, which seems to
very useful for numerical calculations. Remind that in traditional
approached the $1D$ scattering problem Eq. (\ref{34})-(\ref{35})
is constructed after very grate volume of grid computations of
$1D$ Schr\"odinger problem along the reaction coordinate.

The main theoretical advantage is that the wavefunction and all
the body-fixed $\emph{\textbf{S}}$-matrix elements are determined
from the solution of standard coupled differential equations in
only one variable $u$ (the scattering coordinate) simultaneously.
The body-fixed $\emph{\textbf{S}}$-matrix  is used to determine
all the differential and integral  state-to-state reactive
scattering cross sections of the system (see formulas (\ref{30})
and (\ref{31})\,).

It is obvious  that on the basis of developed
$\textbf{\emph{S}}$-matrix representations the maximally-possible
effective parallel algorithms for direct numerical simulation of
$3D$ reactive quantum scattering problem may be elaborated.
Particularly numerical solution of both  systems of inelastic
scattering equations (\ref{39}) or (\ref{41})-(\ref{43}) and
(\ref{60}) can be realized with the help of the
$\textbf{\emph{R}}$-matrix propagation method simultaneously
yields the full wavefunction and all $\textbf{\emph{S}}$-matrix
elements without further calculations.

Finally, we would like to note that the usage  of NCC system in
quantum multichannel scattering theory will permit to carry out
similar type reduction for any amount of atoms. This will
essentially simplify the calculation.

\section{Acknowledgments}

This work partially was  supported by INTAS Grant No. 03-51-4000,
Armenian Science Research Council and Swedish Science Research
Council. AG also thanks ISTC grant N-823, Bristol and G\"{o}teborg
Universities for support of his visit.
\appendix
\section{Choice of 3D NCC system}
Two collinear collision configurations between free particle $A$
and bound pair $(BC)$ schematically may be represented on the
plane of Jacobi coordinates $(R,r)$ in the following case (see
FIG. 4):
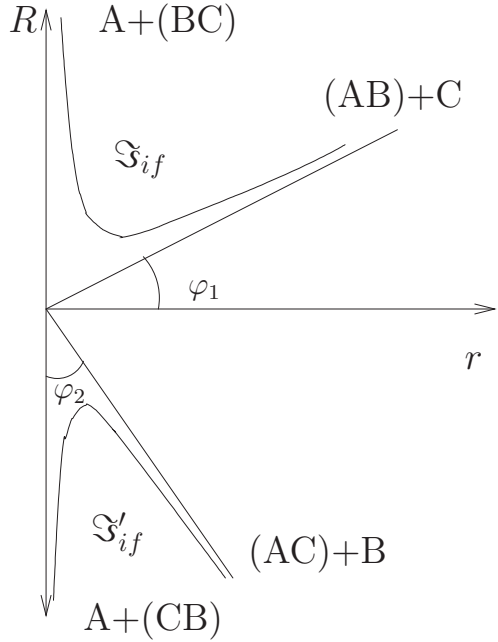
\begin{figure}
\begin{center}
\setlength{\unitlength}{0.00087489in}
\begingroup\makeatletter\ifx\SetFigFont\undefined%
\gdef\SetFigFont#1#2#3#4#5{%
  \reset@font\fontsize{#1}{#2pt}%
  \fontfamily{#3}\fontseries{#4}\fontshape{#5}%
  \selectfont}%
\fi\endgroup%
{\renewcommand{\dashlinestretch}{30}
\begin{picture}(2960,3816)(0,-10)
\put(540.000,2056.500){\arc{732.547}{5.5412}{6.4685}}
\put(292.500,1741.500){\arc{342.710}{0.4049}{1.9757}}
\path(255.000,3669.000)(225.000,3789.000)(195.000,3669.000)
\path(225,3789)(225,144)
\path(195.000,264.000)(225.000,144.000)(255.000,264.000)
\path(225,1989)(2925,1989)
\path(2805.000,1959.000)(2925.000,1989.000)(2805.000,2019.000)
\path(225,1989)(2340,3069)
\path(225,1989)(1350,369)
\path(315,3744)(315,3743)(315,3739)
    (316,3733)(316,3724)(317,3712)
    (318,3696)(319,3677)(321,3655)
    (323,3631)(324,3605)(326,3577)
    (328,3547)(330,3517)(333,3486)
    (335,3454)(338,3421)(340,3388)
    (343,3352)(346,3316)(349,3278)
    (353,3239)(356,3199)(360,3159)
    (365,3109)(369,3064)(373,3028)
    (376,3000)(378,2979)(380,2965)
    (381,2955)(382,2949)(383,2945)
    (383,2941)(384,2937)(385,2929)
    (387,2919)(389,2904)(392,2883)
    (396,2858)(400,2829)(405,2799)
    (412,2758)(418,2726)(422,2705)
    (425,2694)(427,2688)(428,2686)
    (428,2685)(430,2682)(433,2674)
    (437,2661)(443,2642)(450,2619)
    (458,2595)(464,2578)(467,2570)
    (466,2567)(465,2567)(464,2567)
    (465,2565)(469,2558)(479,2546)
    (495,2529)(511,2514)(525,2502)
    (536,2492)(544,2486)(550,2480)
    (555,2476)(560,2473)(566,2468)
    (576,2463)(590,2456)(608,2447)
    (630,2439)(651,2433)(668,2428)
    (678,2425)(682,2423)(681,2422)
    (679,2420)(677,2419)(678,2419)
    (685,2421)(701,2424)(728,2430)
    (765,2439)(793,2447)(821,2456)
    (845,2463)(865,2470)(880,2475)
    (892,2480)(900,2483)(906,2485)
    (911,2488)(917,2490)(924,2494)
    (935,2498)(950,2504)(971,2513)
    (998,2524)(1034,2538)(1077,2555)
    (1125,2574)(1163,2589)(1200,2604)
    (1234,2618)(1265,2630)(1292,2641)
    (1315,2651)(1335,2659)(1351,2665)
    (1364,2671)(1376,2675)(1386,2679)
    (1395,2683)(1404,2686)(1414,2690)
    (1426,2695)(1439,2701)(1455,2708)
    (1475,2716)(1498,2726)(1525,2737)
    (1556,2751)(1590,2766)(1627,2782)
    (1665,2799)(1709,2819)(1749,2837)
    (1785,2855)(1818,2870)(1846,2884)
    (1872,2897)(1895,2908)(1916,2919)
    (1935,2929)(1952,2938)(1968,2947)
    (1982,2954)(1994,2961)(2004,2967)
    (2012,2972)(2018,2975)(2022,2977)
    (2024,2978)(2025,2979)
\path(270,234)(270,235)(270,239)
    (270,245)(271,254)(271,267)
    (272,284)(273,305)(274,329)
    (275,357)(277,387)(278,420)
    (280,455)(281,491)(283,529)
    (285,567)(287,605)(289,644)
    (291,683)(293,723)(295,762)
    (298,803)(300,843)(303,884)
    (306,925)(309,966)(312,1006)
    (315,1044)(321,1111)(327,1161)
    (331,1195)(334,1213)(336,1221)
    (337,1221)(338,1216)(338,1211)
    (339,1207)(341,1208)(344,1214)
    (348,1227)(354,1246)(360,1269)
    (369,1297)(375,1315)(379,1325)
    (381,1329)(383,1329)(384,1329)
    (386,1330)(390,1336)(396,1346)
    (405,1359)(418,1377)(425,1386)
    (428,1389)(430,1391)(437,1396)
    (450,1404)(457,1407)(463,1409)
    (467,1411)(468,1413)(469,1414)
    (469,1415)(469,1416)(470,1416)
    (472,1416)(477,1414)(485,1410)
    (495,1404)(508,1394)(519,1386)
    (525,1381)(528,1379)(529,1378)
    (530,1376)(534,1372)(544,1361)
    (561,1342)(585,1314)(602,1294)
    (617,1276)(629,1262)(637,1253)
    (640,1249)(641,1249)(640,1251)
    (638,1254)(636,1256)(640,1251)
    (649,1240)(665,1219)(690,1187)
    (723,1143)(765,1089)(790,1057)
    (815,1023)(841,989)(867,954)
    (893,920)(919,886)(944,852)
    (969,819)(994,786)(1018,753)
    (1042,721)(1066,689)(1090,657)
    (1113,626)(1136,595)(1159,566)
    (1180,537)(1200,510)(1219,484)
    (1237,461)(1253,439)(1266,421)
    (1278,406)(1287,393)(1294,384)
    (1299,377)(1302,372)(1304,370)(1305,369)
\put(540,3654){\makebox(0,0)[lb]{\smash{{{\SetFigFont{14}{16.8}{\rmdefault}{\mddefault}{\updefault}A+(BC)}}}}}
\put(1890,3204){\makebox(0,0)[lb]{\smash{{{\SetFigFont{14}{16.8}{\rmdefault}{\mddefault}{\updefault}(AB)+C}}}}}
\put(630,2844){\makebox(0,0)[lb]{\smash{{{\SetFigFont{14}{16.8}{\rmdefault}{\mddefault}{\updefault}$\Im_{if}$}}}}}
\put(450,54){\makebox(0,0)[lb]{\smash{{{\SetFigFont{14}{16.8}{\rmdefault}{\mddefault}{\updefault}A+(CB)}}}}}
\put(1080,2079){\makebox(0,0)[lb]{\smash{{{\SetFigFont{12}{14.4}{\rmdefault}{\mddefault}{\updefault}$\varphi_1$}}}}}
\put(0,3654){\makebox(0,0)[lb]{\smash{{{\SetFigFont{14}{16.8}{\rmdefault}{\mddefault}{\updefault}$R$}}}}}
\put(2745,1629){\makebox(0,0)[lb]{\smash{{{\SetFigFont{14}{16.8}{\rmdefault}{\mddefault}{\updefault}$r$}}}}}
\put(495,594){\makebox(0,0)[lb]{\smash{{{\SetFigFont{14}{16.8}{\rmdefault}{\mddefault}{\updefault}$\Im_{if}'$}}}}}
\put(1440,459){\makebox(0,0)[lb]{\smash{{{\SetFigFont{14}{16.8}{\rmdefault}{\mddefault}{\updefault}(AC)+B}}}}}
\put(270,1450){\makebox(0,0)[lb]{\smash{{{\SetFigFont{12}{14.4}{\rmdefault}{\mddefault}{\updefault}$\varphi_2$}}}}}
\end{picture}
}
\end{center}
\caption{\emph{The curve of reaction coordinates for two different
exchange reactions on a collinear collisions plane $(R,r)$. It is
obvious that in this plane there are not interferences between two
exchange reaction channels. Recall that angles $\varphi_1$ and
$\varphi_2$ are given by formulas
$cot\varphi_{1}=\bigl[{m_Am_C}/{m_BM}\bigr]^{1/2}$ and
$cot\varphi_{2}=\bigl[{m_Am_B}/{m_CM}\bigr]^{1/2}$. } }
\end{figure}
\begin{figure}
\begin{center}
\includegraphics[width=0.22\textwidth]{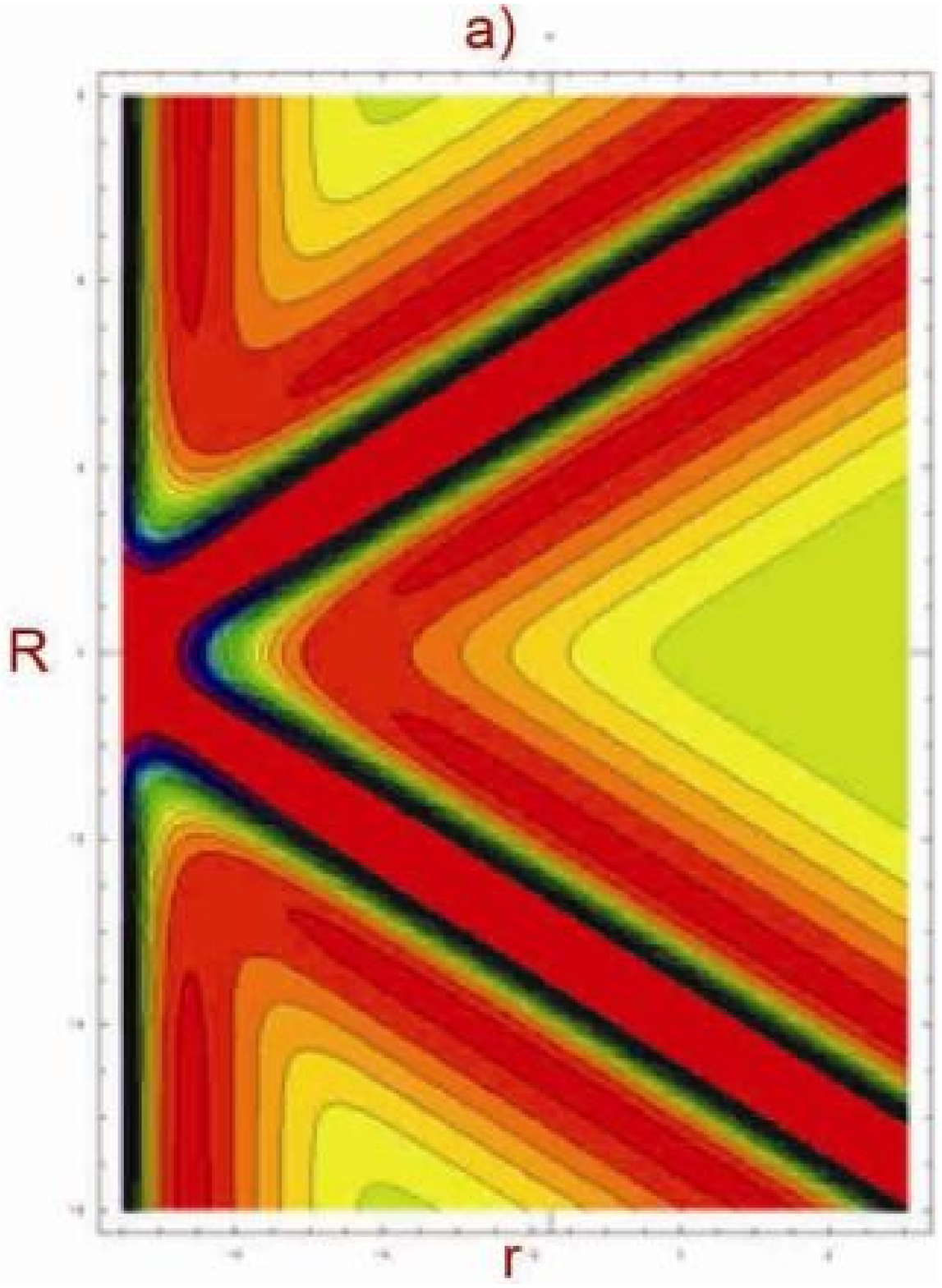}
\includegraphics[width=0.22\textwidth]{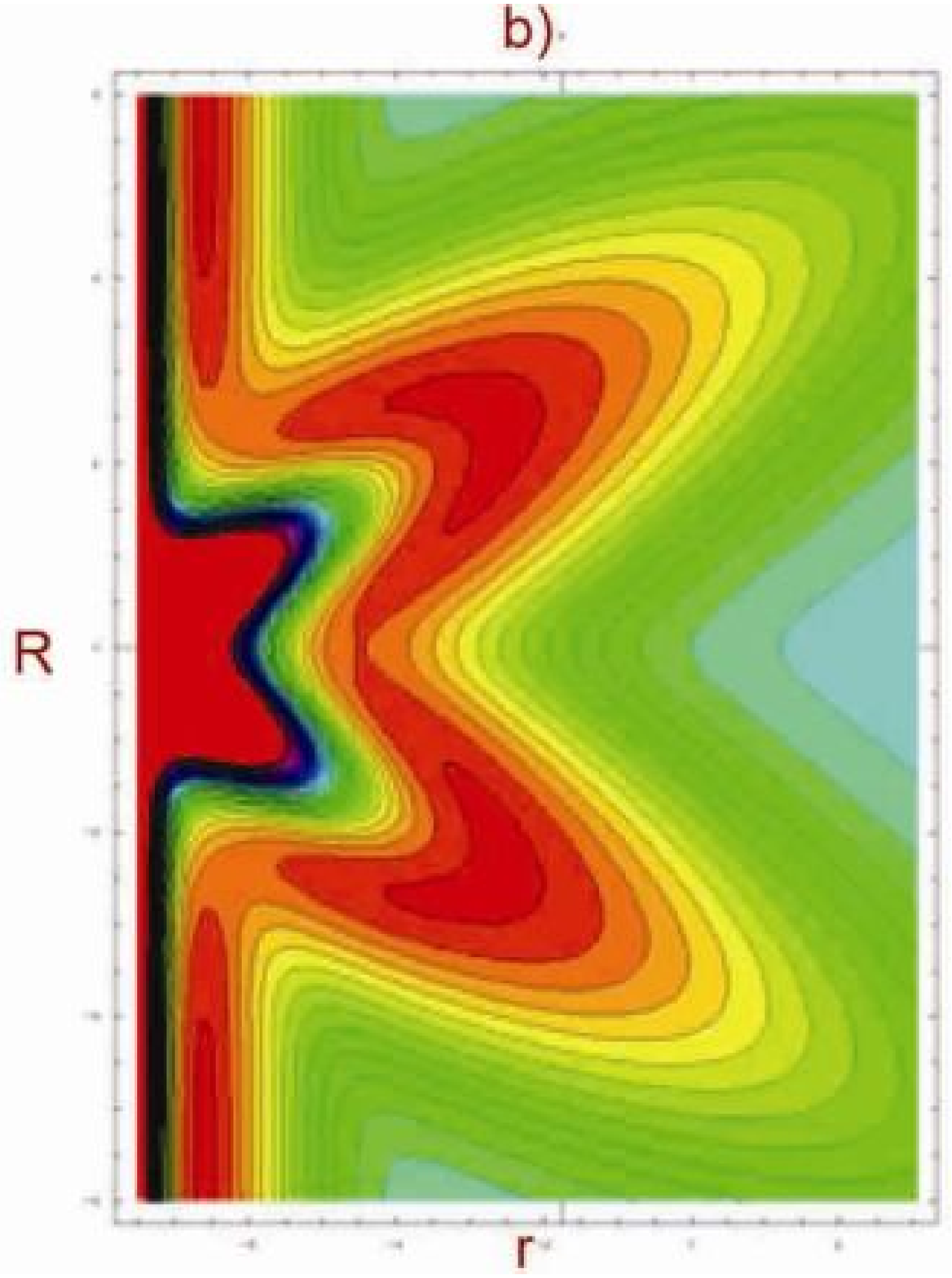}
\includegraphics[width=0.22\textwidth]{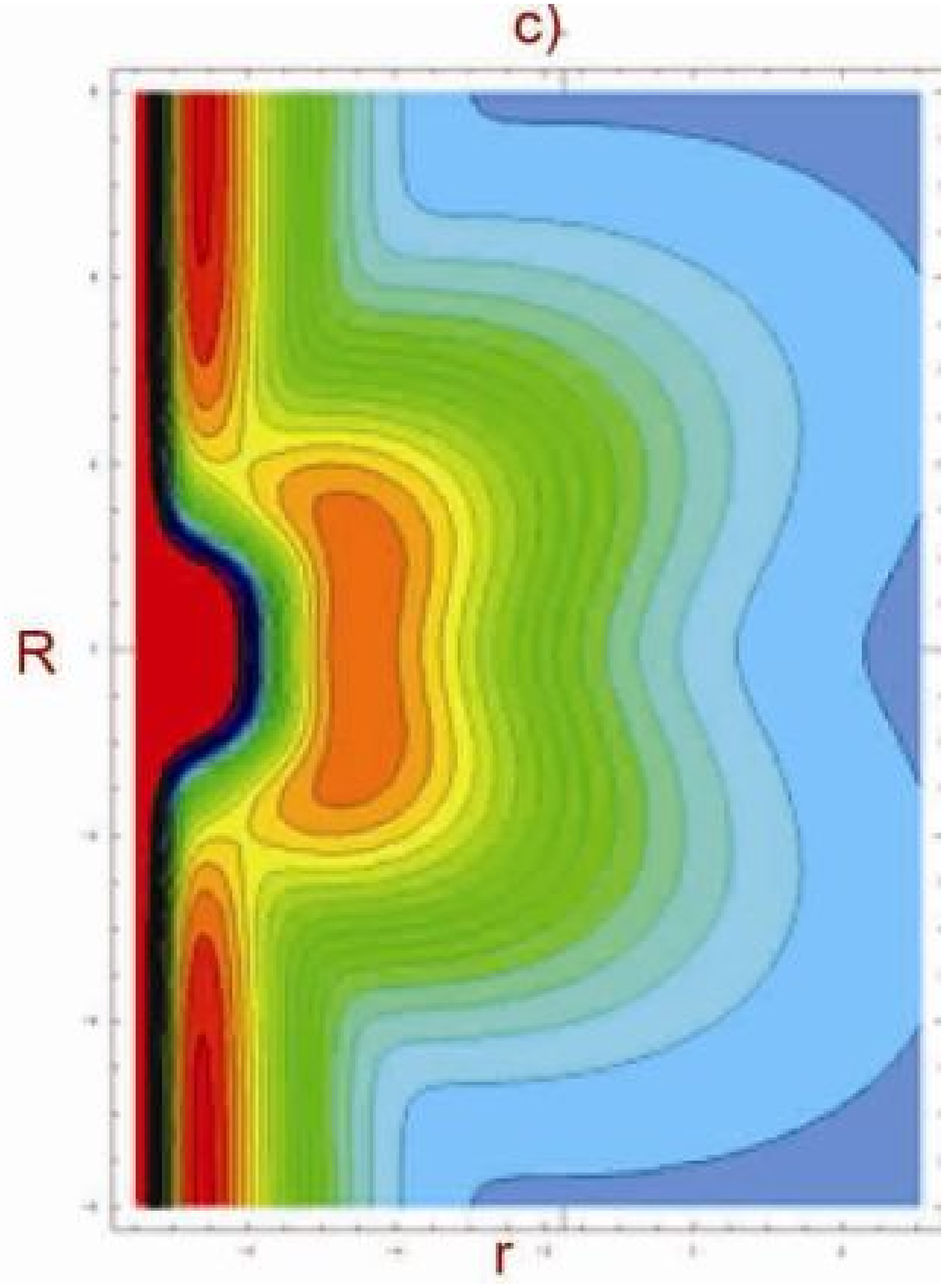}
\includegraphics[width=0.22\textwidth]{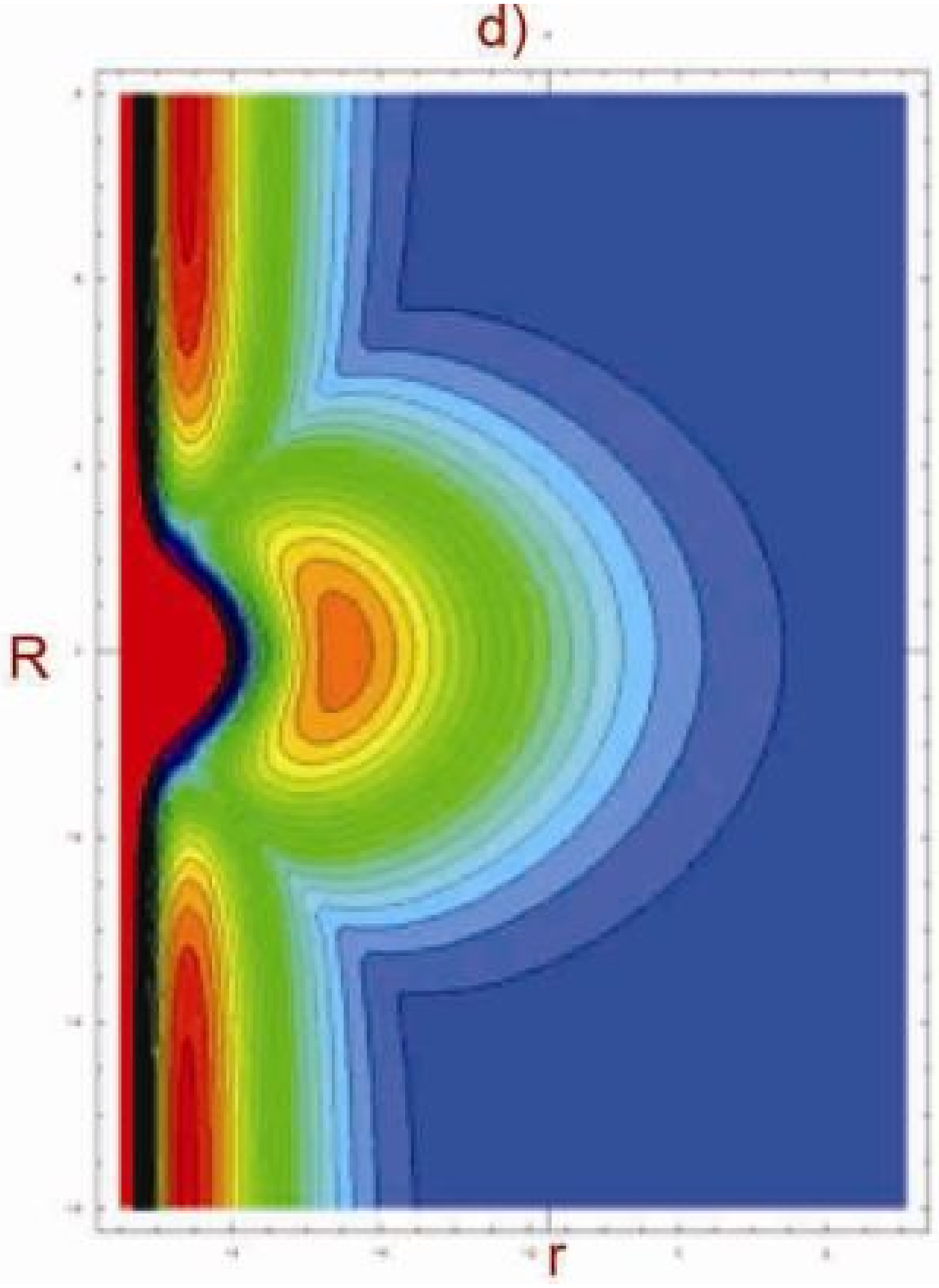}
\end{center}
\caption{\emph{Energy surfaces of reactive system $H+H_2$ in the
plane of Jacobi coordinates  $(R,r)$ for different angles
$\theta=0^0,\,30^0,\,60^0,\,90^0$. As  it is evidently in the
plane a) two reaction channels are opened while on the planes
b),\,c),\,d)  only excitation channels are opened. Note that it is
a universal property which follows from kinematic geometry of
scattering and doesn't depend on what $A, \,B$ and $C$ particles
are interacted.}}
\end{figure}
The smooth curves (coordinate reactions) $\Im_{if}$ and
${\Im'}_{if}$ in figure connect  correspondingly reagent and
product asymptotic subspaces  in the collinear collisions $A+(BC)$
and $A+(CB)$. It is obvious that in this case the different
reaction channels are isolated and there is no interference
between them. The illustration of this fact is shown on the
example of \emph{ab initio} PES, termed LSTH, type potential
energy surface for reactive system $H+H_2$
 (see FIG. 5a)  \cite{Sieg,Truhl}. When the Jacobi angle is fixed
$\theta\in(0,\pi)$, going along the curve $\Im_{if}$ or
${\Im'}_{if}$ we can't come to mentioned products subspaces. In
other words the exchange reaction channels on the plane $(R,r)$ in
this case are closed and this is an universal kinematic property
of scattering and doesn't depend from particles sort (see FIG.
5b-5d).
\begin{figure}
\begin{center}
\includegraphics[width=0.45\textwidth]{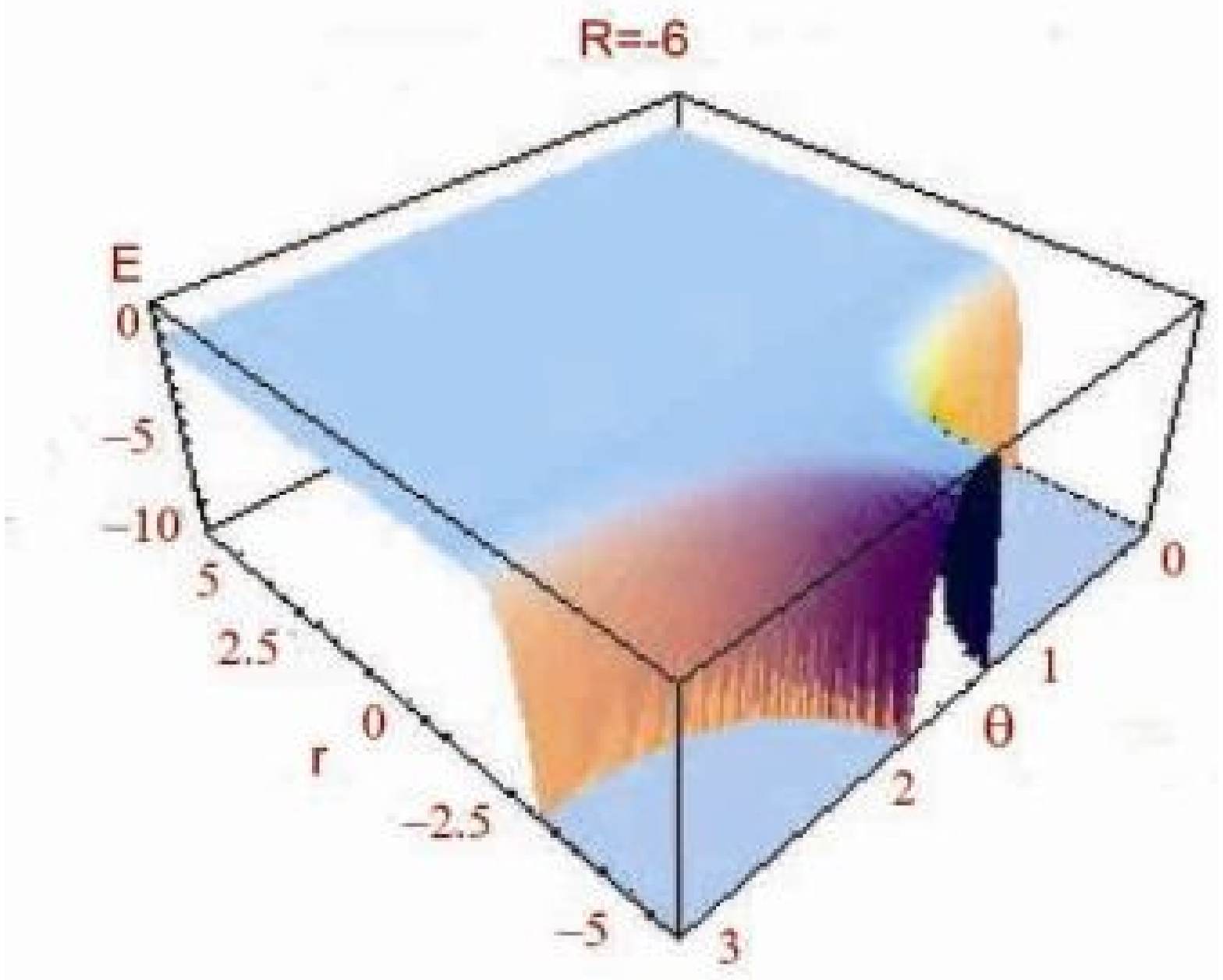}\qquad
\includegraphics[width=0.45\textwidth]{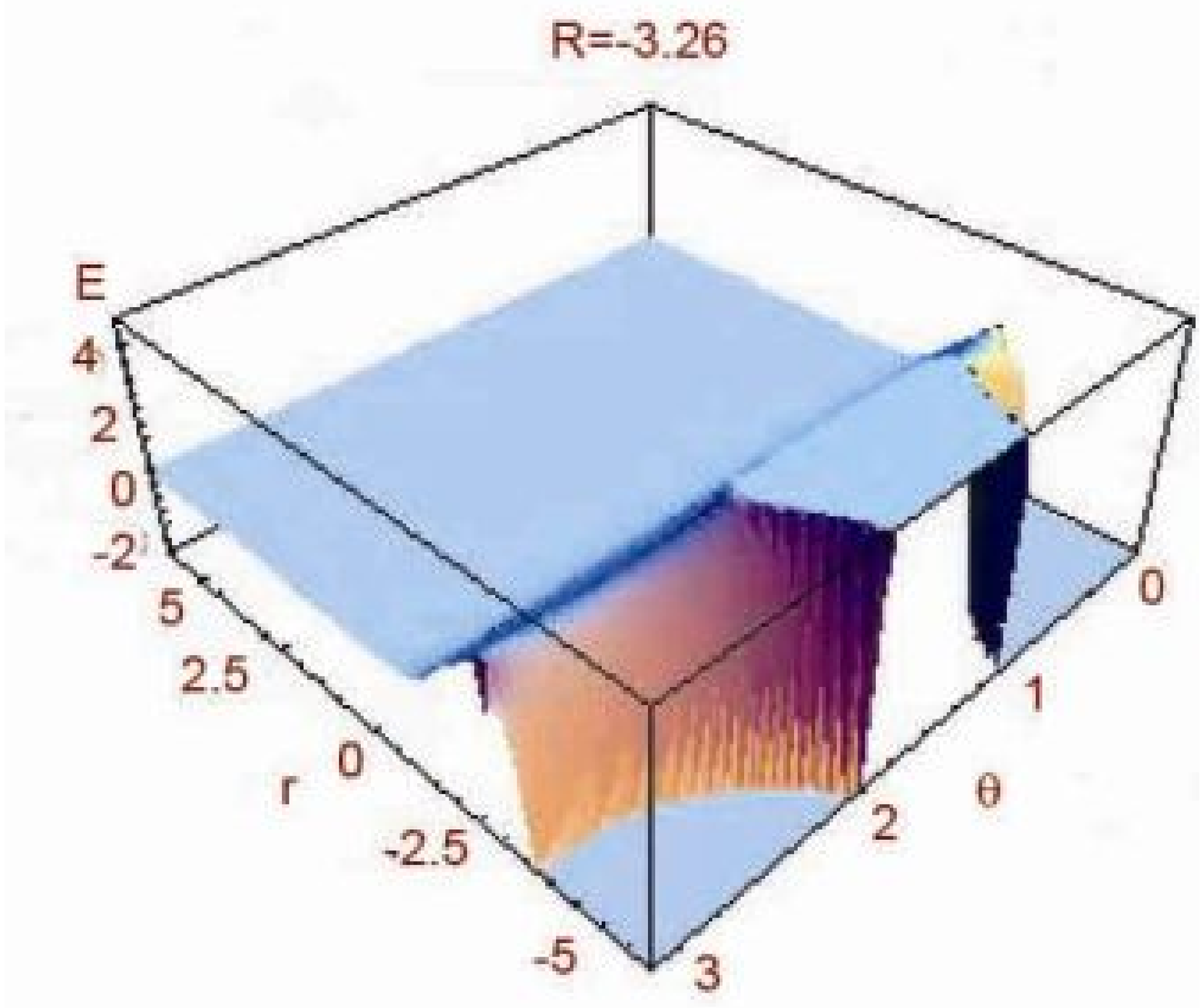}
\includegraphics[width=0.45\textwidth]{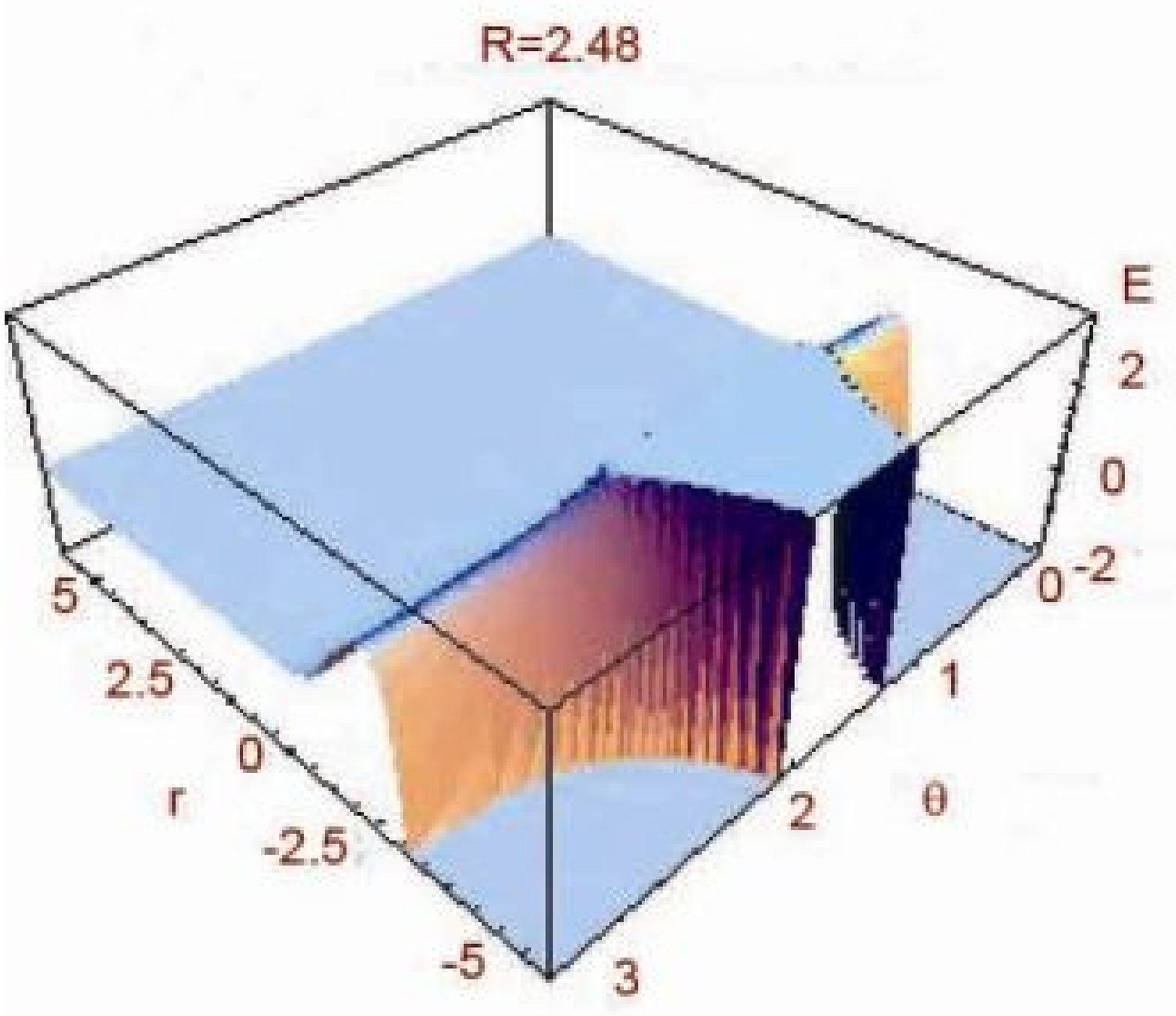}\qquad
\includegraphics[width=0.45\textwidth]{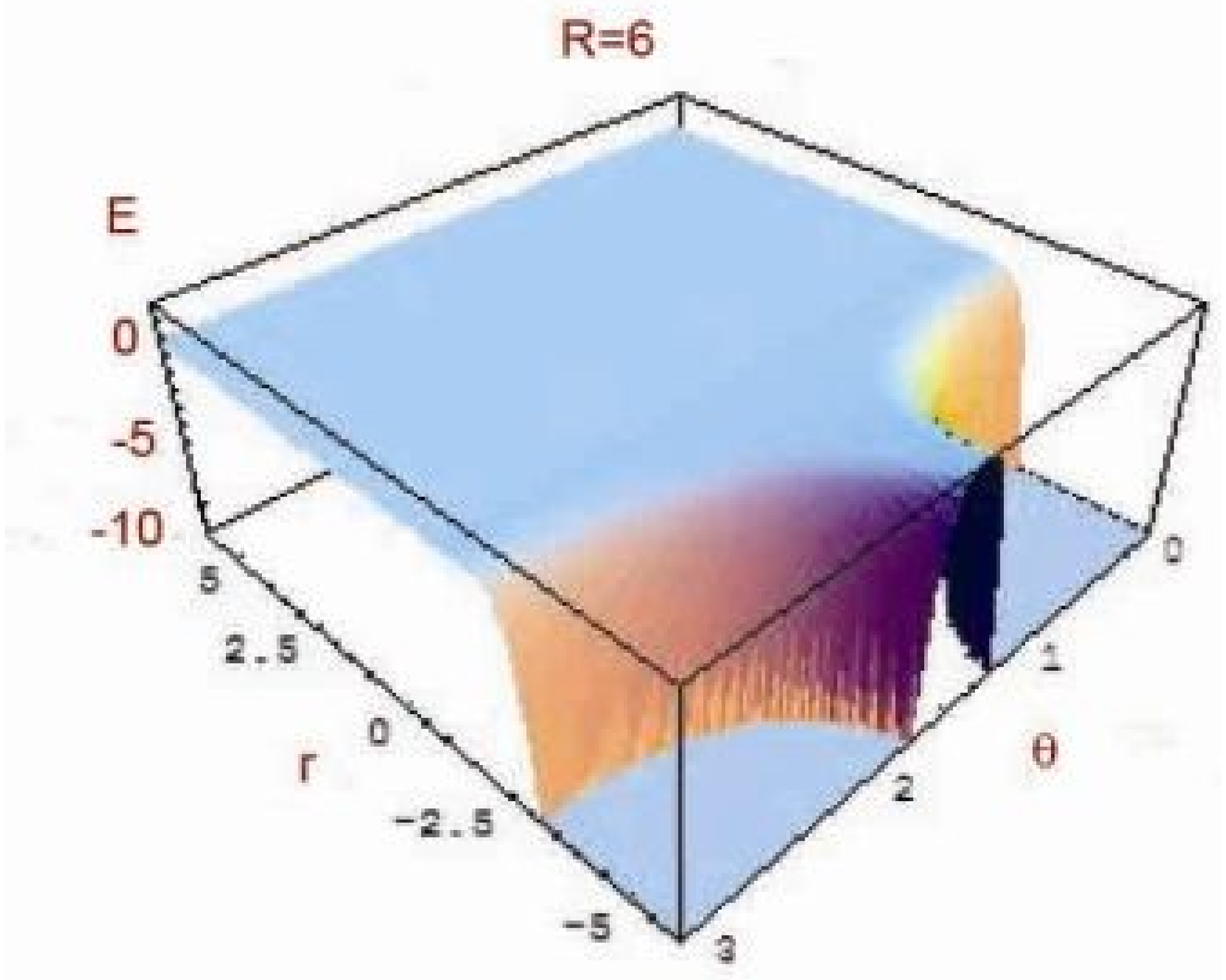}
\end{center}
\caption{\emph{Energetic surfaces of reactive system $H+H_2$ in the
 coordinates plane $(r,\theta)$ for different fixing coordinates
$R$. It is evident from pictures in the strong interaction region
(the region where interactions between all particles are
essentially) that there are two potential holes one around the
angle $\theta=0$ (or collinear configuration $A+(BC)$) and  the
second around the angle $\theta=\pi$ (or collinear configuration
$A+(CB)$). It is obviously that during the evolution the three
body system with big probabilities can fall within  first or
second hole. }}
\end{figure}

\begin{figure}
\begin{center}
\includegraphics[width=0.45\textwidth]{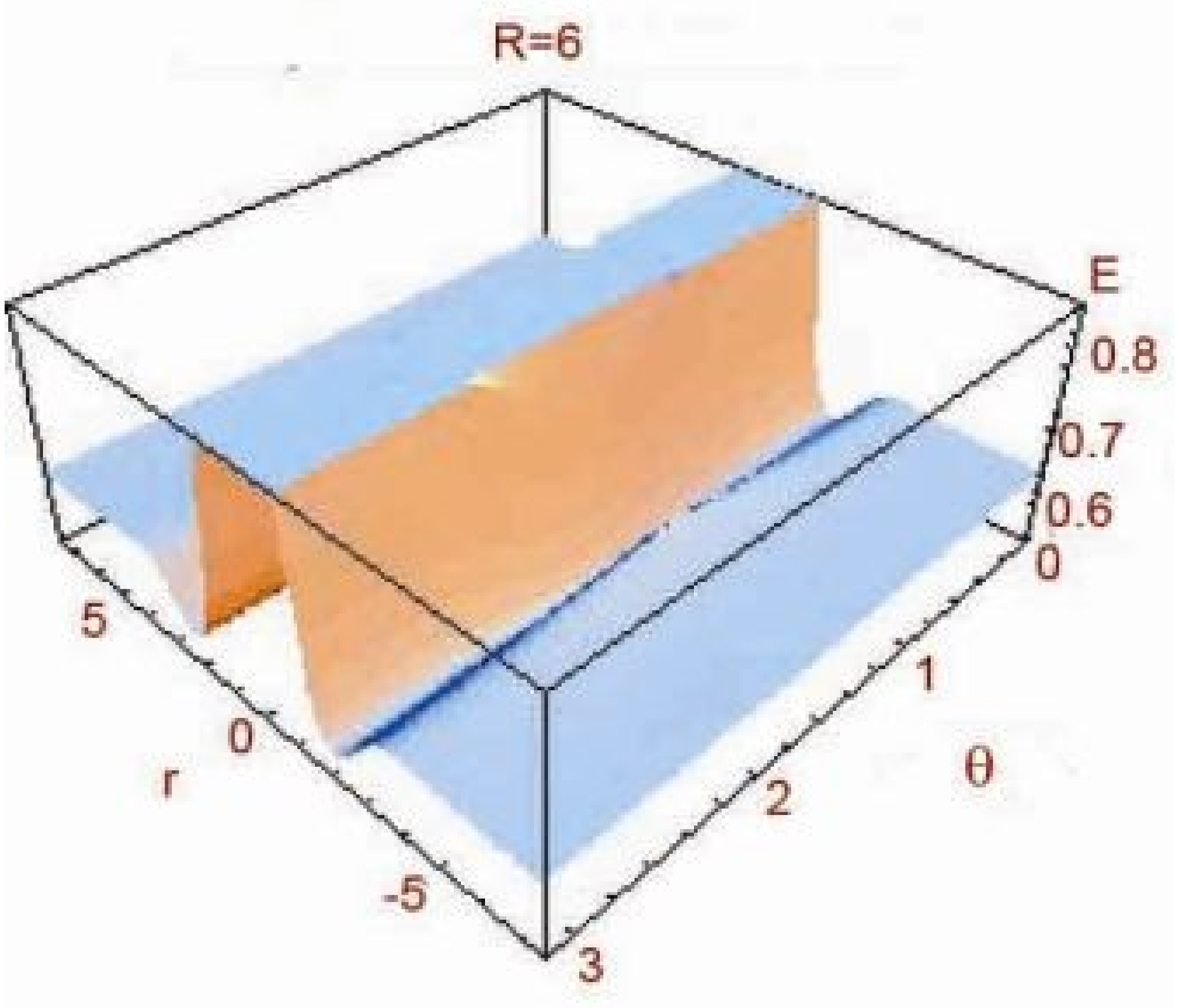}\qquad
\includegraphics[width=0.45\textwidth]{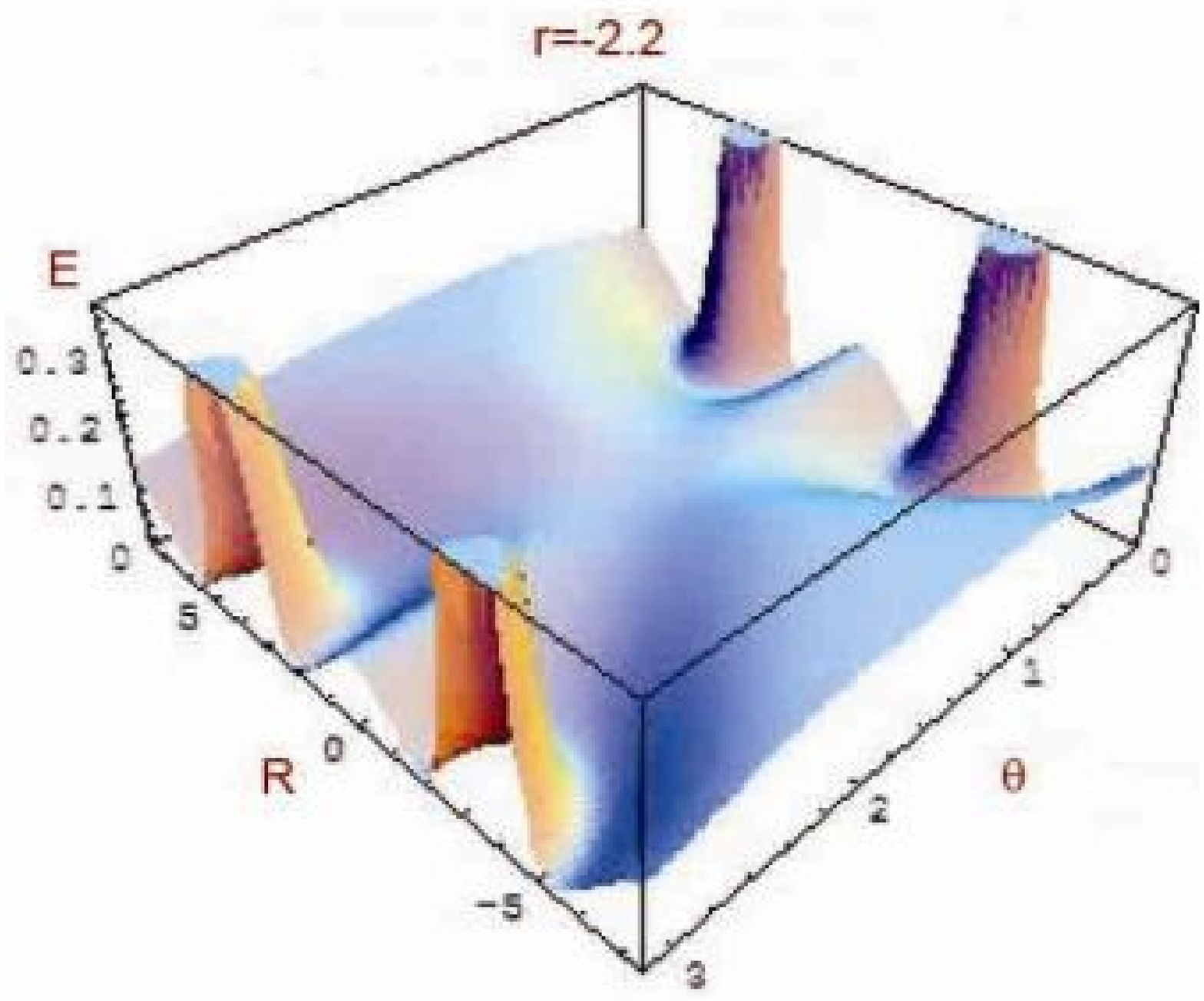}
\\
\includegraphics[width=0.45\textwidth]{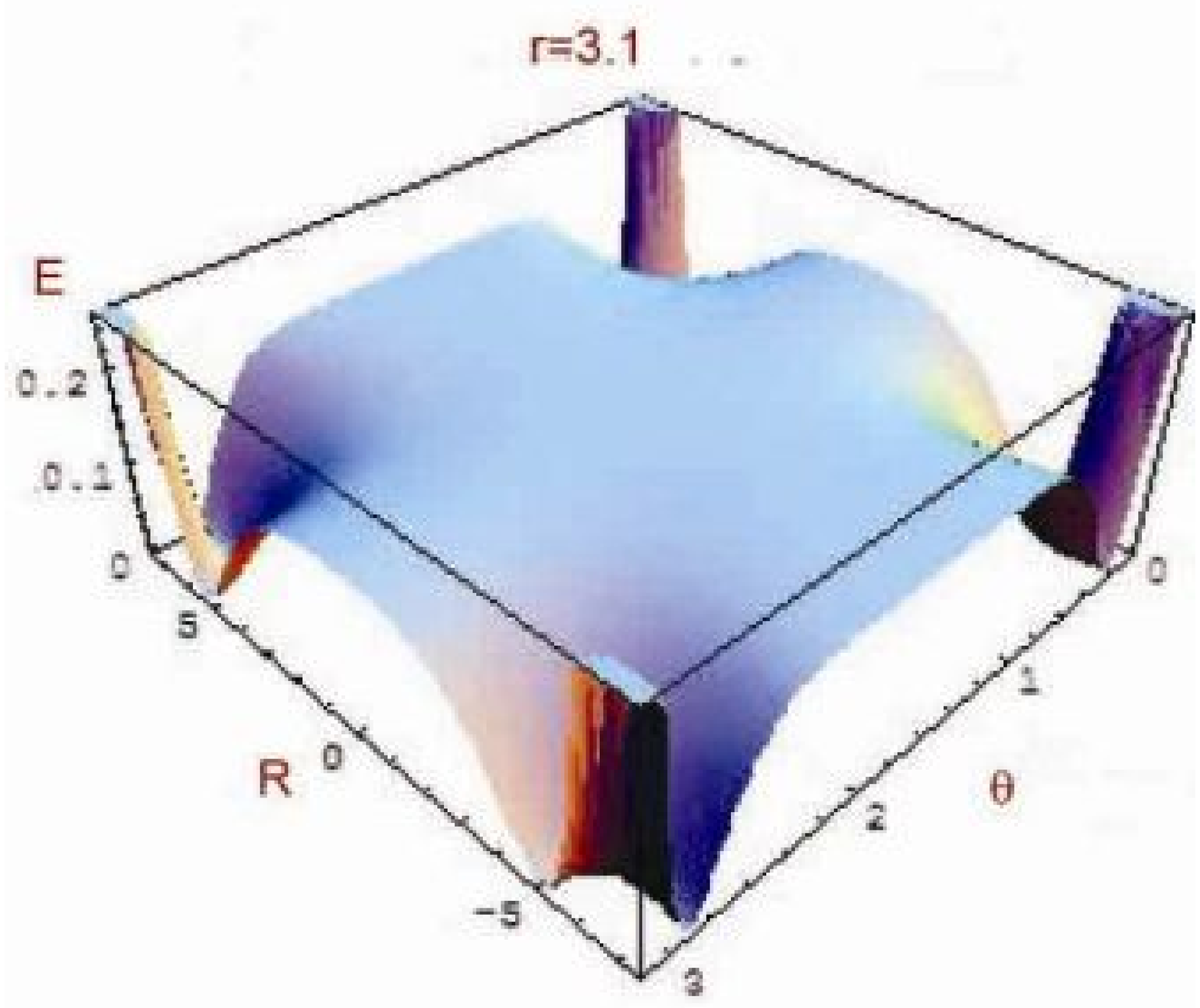}\qquad
\includegraphics[width=0.45\textwidth]{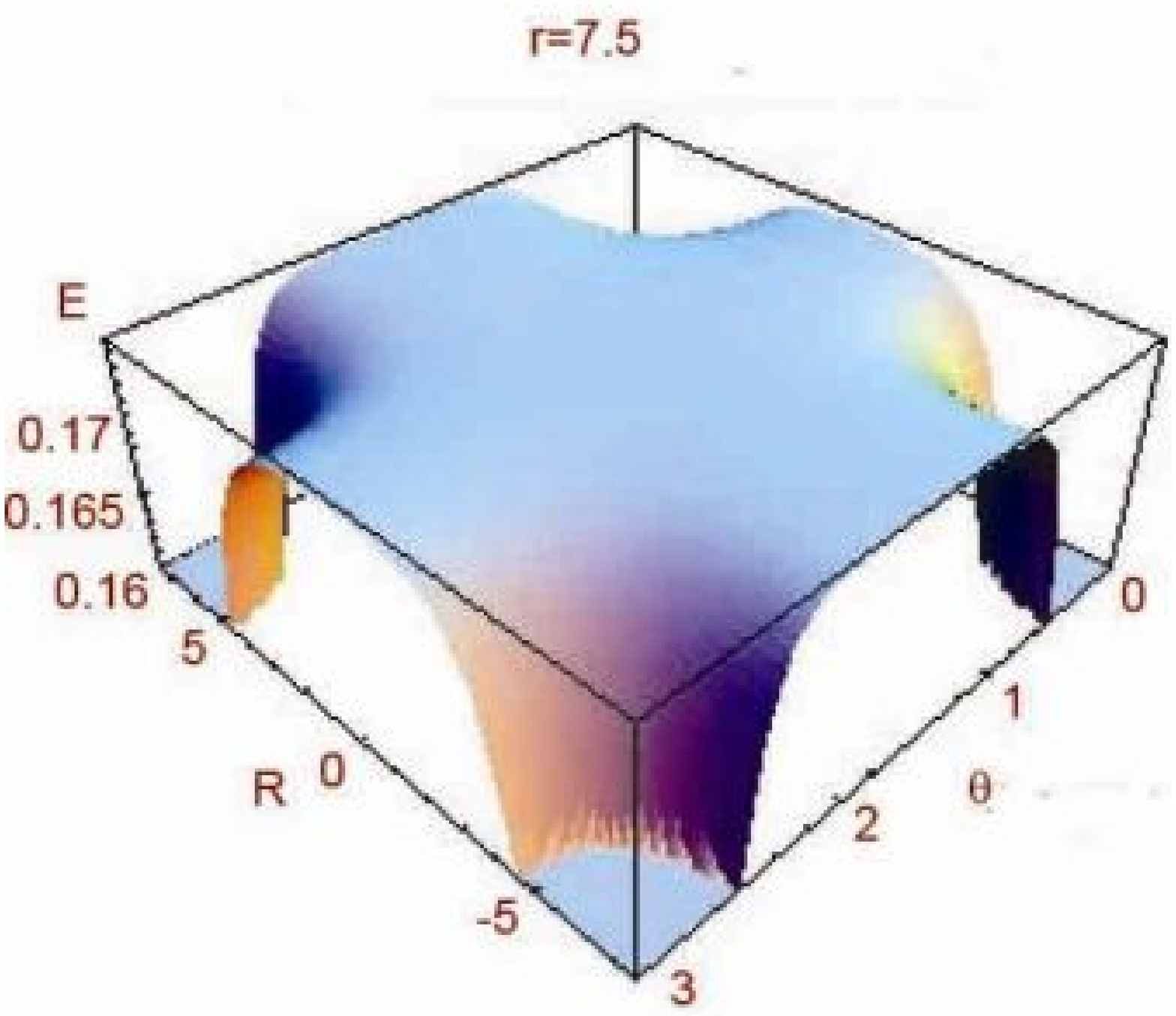}
\end{center}
\caption{Energetic surfaces of reactive system $H+H_2$ in the
 coordinates plane $(R,\theta)$ for different fixing coordinates
$r$. On these energetic surfaces are generated gradients which
push on the three body system to collinear configurations.}
\end{figure}
However, in this case the reaction goes out of mentioned plane in
$3D$ space (see FIG. 6a-6d and FIG. 7a-7d). As kinematic analysis
shows, the $3D$ region, where the probability of reactive
scattering processes is concentrated, is essential around the
collinear collision plane $(R,r)$. Moreover, this region is
compressed to the plan $(R,r)$ as system is evolving along
coordinate reaction curve $\Im_{if}$ (or ${\Im'}_{if}$) to the
asymptotic subspace of products. Recall that all topological
properties of reaction surface, which are in Jacobi coordinates
system, are conserved in $3D$ NCC system too.

 So, the $3D$ NCC system, which
is connected with the  reaction coordinate curves $\Im_{if}$ or
${\Im'}_{if}$, describes uniquely $3D$ part of configuration
space, where multichannel scattering process is concentrated.

\section{Mass-scaled distances between three-body}

For right introducing of NCC system it is necessary to connect the
curve of reaction coordinate $\Im_{if}$ in the plan of $(q_0,q_1)$
with  two equilibrium distances of diatoms of $(in)$ and $(out)$
asymptotic channels. Let us consider the distant between $A$ and
$B$ particles in the $(in)$ Jacobi coordinates (see FIG. 1):
\begin{equation}\label{B1}
R_{AB}=\sqrt{R^2_\alpha-2\rho R_\alpha r_\alpha\cos\theta_\alpha
+(\rho r_\alpha)^2},\quad \rho=\frac{m_C}{m_b+m_C},
\end{equation}
where indexes which note channels are admitted. After simple
transformations taking into account (\ref{02}) and (\ref{03}) from
(\ref{B1}) we can find:
\begin{equation}\label{B2}
R_{AB}=\sqrt{\lambda^{-2}q_0^2-2\rho q_0q_1\cos\theta
+\lambda^{2}(\rho q_1)^2}=\lambda^{-1}\sqrt{q_0^2-2b
q_0q_1\cos\theta +(b q_1)^2}=\lambda^{-1}q_2,
\end{equation}
where $b=\rho\lambda^2=\cot\varphi$, and $q_2$ is mass-scaled
distance between $A$ and $B$ particles. Now we can investigate the
asymptotic behavior of mass-scaled distance $q_2$. From Eq.
(\ref{B2}) with taking into account (\ref{05})  in the $(in)$
channel (in the limit of $q_1^c\to q_{eq}^-$) may be find:
\begin{equation}
q_2\bigl|_{\Im_{if}}-q_0^c=\bigl(q_2^c-q_0^c\bigr)\bigl|_{q_1^c\to
q_{eq}^-}=-bq_{eq}^-\cos\theta. \label{B3}
\end{equation}
From which is following that curve $\Im_{if}$ in the reactant
$(in)$ channel comes to equilibrium diatom $(BC)$ distance
$q_{eq}^-$.

 Because  Jacobi scattering angle which is defined in $(in)$
channel in the $(out)$ channel  is limited to zero ( $\theta\to
0$) from (\ref{B2}) and (\ref{05}) can be find:
\begin{equation}
q_2^c\bigl|_{q_1^c\to \infty}=q_0^c-bq_1^c=q_{eq}^+,\label{B4}
\end{equation}
which is mean that in product $(out)$ channel curve $\Im_{if}$
coming to equilibrium distance  $q_{eq}^+$ of  $(AB)$ diatom.

 So, the definition of curve of \emph{reaction coordinate }
$\Im_{if}$ which is defined by formulas (\ref{05}) and (\ref{06})
 satisfies the aforementioned asymptotic conditions.

 Completely note that definition of curve of reaction coordinate
 $\Im_{if}'$ for exchange reaction $A+(CB)\to (AC)+B$ we can
 find from (\ref{05}) putting instead of constants $b$
 and $q_{eq}^+$  constants $(1-b)$ and $\widetilde{q}_{eq}^+$
 correspondingly,
 where $\widetilde{q}_{eq}^+$ is an equilibrium distance between
 pair $(AC)$ in the asymptotic channel ($q_1^c\to+\infty$).

As for the dissociation reaction $A+ (BC)\to A+B+C$ it can be
described by term of both NCC systems. Only in this case we must
remember that Jacobi angle does not limit to zero and can have any
value.

\section{Calculation of metric tensor in curvilinear reaction coordinates}

 Let us calculate the transformation between Jacobi scaled and NCC
 coordinates systems in the plan $(q_0,q_1)$. As well
 known covariant metric
  tensor defined by:
\begin{equation}\label{C1}
\gamma_{ij}=\sum_{k=0}^{1}\frac{\partial{q_k}}{\partial{{x}^i}}
\frac{\partial{q_k}}{\partial{x}^j},\qquad\{{x}^j\}
=(x^0,x^1)\equiv(u,v),\quad i,j,k=0,1.
\end{equation}
In order to do this, first we calculate the non-diagonal terms.
Using (\ref{05}) and (\ref{06}), as well as the fact that $x^0$
and $x^1$ are orthogonal, one can obtain from (\ref{C1}) the
following expression:
\begin{eqnarray}\label{C2}
\gamma_{01}=\gamma_{10}=-\biggl(\frac{dq_0^c}{d{u}}
-v\cos{\phi(u)}\frac{d\phi}{du}\biggr)\sin{\phi}+\biggl(\frac{dq_1^c}{d{u}}
-v\sin{\phi(u)}\frac{d\phi}{du}\biggr)\cos{\phi}
\nonumber\\=-\frac{dq_0^c}{d{u}}
\sin{\phi(u)}+\frac{dq_1^c}{d{u}}\cos{\phi(u)}.
\end{eqnarray}

Requiring orthogonality of the NCC we set $\gamma_{01}=
\gamma_{10}=0$, giving:
\begin{equation}\label{C3}
\frac{dq_0^c}{dq_1^c}=\cot{\phi{(u)}},
\end{equation}
on the curve $\Im_{ij}$. Note that the function $\phi{(u)}$ has a
simple geometrical meaning: it corresponds to the angle between
the tangential to the reaction coordinate and $q_1$. From
(\ref{C3}) it is easy to get:
\begin{equation}
\label{C4}
\sin\phi(u)=\frac{1}{\sqrt{1+\bigl({dq_0^c}\bigl/{dq_1^c}\bigr)^2}},\qquad
\cos\phi(u)=\frac{{dq_0^c}\bigl/{dq_1^c}}
{\sqrt{1+\bigl({dq_0^c}\bigl/{dq_1^c}\bigr)^2}}.
\end{equation}
After derivation $\sin\phi(u)$ by variable $u$ with taking into
account (\ref{C4}) one can write:
\begin{equation}\label{C5}
\bigl(\sin\phi(u)\bigr)_{;\,u}=\cos\phi(u)\,\frac{d\phi}{du}=-
\frac{{dq_0^c}\bigl/{dq_1^c}}{\Bigl\{{1+\bigl({dq_0^c}\bigl/{dq_1^c}
\bigr)^2\Bigr\}^{3/2}}}\frac{d}{du}\Bigl(\frac{dq_0^c}{dq_1^c}\Bigr).
\end{equation}
Using equations (\ref{05}) and (\ref{06}) from (\ref{C5}) may be
find:
\begin{equation}\label{C6}
\bigl(\sin\phi(u)\bigr)_{;\,u}=
-\frac{2a}{\bigl(q_1^c-q_{eq}^-\bigr)^3}\frac{b-a/\bigl(q_1^c-q_{eq}^-\bigr)^2}
{b+a/\bigl(q_1^c-q_{eq}^-\bigr)^2}{\biggl\{{1+\Bigl[b-a/{\bigl(q_1^c-q_{eq}^-\bigr)^2}
\Bigr]^2\biggr\}^{-3/2}}},
\end{equation}
where symbol $()_{;s}$ is denoted derivation.

The expression for function $\bigl(\sin\phi(u)\bigr)_{;u}$
correspondingly have a form:
\begin{eqnarray}
\label{C7}
\bigl(\cos\phi(u)\bigr)_{;\,u}=-\frac{2a}{\bigl(q_1^c-q_{eq}^-\bigr)^3}
\frac{1}
{b+a/\bigl(q_1^c-q_{eq}^-\bigr)^2}{\biggl\{{1+\Bigl[b-a/{\bigl(q_1^c-q_{eq}^-\bigr)^2}
\Bigr]^2\biggr\}^{-3/2}}}.
\end{eqnarray}

 For matrix element $\gamma_{00}$ from (\ref{C1}) may be find expression:
\begin{eqnarray}\label{C8}
\gamma_{00}= \biggl(\frac{\partial{q_0}}{\partial{u}}\biggr)^2+
\biggl(\frac{\partial{q_1}}{\partial{u}}\biggr)^2=\biggl(\frac{dq_0^c}{d{u}}
-v\cos{\phi(u)}\frac{d\phi}{du}\biggr)^2+\biggl(\frac{dq_1^c}{d{u}}
-v\sin{\phi(u)}\frac{d\phi}{du}\biggr)^2.
\end{eqnarray}

Substituting the derivatives (\ref{C6}) and (\ref{C7}) into
(\ref{C8}) we obtain:
\begin{eqnarray}\label{C9}
\gamma_{00}=\bigl[1+K(u)v\bigr]^2\biggl(\frac{ds}{du}\biggr)^2,
\end{eqnarray}
where $K(u)$ is Gaussian curvature of reaction coordinate curve
$\Im_{if}$ and $s$ is correspondingly the length  along the
$\Im_{if}$:
\begin{eqnarray}\label{C10}
  K(u)=
\frac{2a}{\bigl(q_1^c-q_{eq}^-\bigr)^3}\biggl\{1+
\Bigl[b-a/{\bigl(q_1^c-q_{eq}^-\bigr)^2}
 \Bigr]^2\biggr\}^{-3/2},\nonumber\\
 \biggl(\frac{ds}{du}\biggr)^2=\biggl\{1+\Bigl[b-
a/{\bigl(q_1^c-q_{eq}^-\bigr)^2}\Bigr]^2
 \biggr\}\biggl\{b+a/{\bigl(q_1^c-q_{eq}^-\bigr)^2}\biggr\}^{-2}.
\end{eqnarray}
Note that the Lam$\acute{e}$ coefficient is given by
$\eta=\sqrt{\gamma_{00}}$.

We also note that the elements
$\gamma_{02}=\gamma_{02}=\gamma_{12}=\gamma_{21}$ vanish. Remind
that the second diagonal element $\gamma_{11}$ may be obtained by
using (\ref{05})-(\ref{06}) and (\ref{C1}):
\begin{equation}\label{C11}
\gamma_{11}=\biggl(\frac{\partial{q_0}}{\partial{v}}\biggr)^2+
\biggl(\frac{\partial{q_1}}{\partial{v}}\biggr)^2=1.
\end{equation}
Thus the metric tensor $\gamma_{ij}$ is diagonal. Finally we note
that the covariant tensor is defined as:
\begin{equation}\label{C12}
\gamma^{ij}=B_{ij}\bigl/\gamma, \quad
\gamma=Det\bigl(\gamma_{ii}\bigr)=\gamma_{00}\gamma_{11}
=\gamma_{00},
\end{equation}
where $B_{ij}$ is the algebraic complement to $\gamma_{ij}$.

\section{$\emph{\textbf{S}}$-matrix contraction by reaction
 coordinate (intrinsic time) formalism}

Between full wavefunctions in the different body-fixed systems the
following transformation  may be written (see (\ref{24})\,):
\begin{eqnarray}
\label{D1}
\sum_{\bar{K}}\rho^{-1/2}_\alpha{\Phi}^{(\pm)J}_{nj\bar{K}}
(u_\alpha,v_\alpha,\theta_\alpha)D^J_{\bar{K}K}(\Omega_\alpha)=
\,\sum_{\bar{K}} \rho^{-1/2}_\beta {\Phi}^{(\pm)J}_{nj\bar{K}}
(u_\beta,v_\beta,\theta_\beta)D^J_{\bar{K}K}(\Omega_\beta).
\end{eqnarray}
Next we multiply the equation (\ref{D1}) by the Wigner
$D^{J\ast}_{\bar{K}\bar{K}}(\Omega)$-function
 and integrate, using the
volume element $d\Omega=\sin\zeta_2\,d\zeta_1d\zeta_2d\zeta_3$.
With help of formulas:
$$
\int\,D^{J_1\ast}_{M_1M'_1}(\zeta_1,\zeta_2,\zeta_3)D^{J_2}_{M_2M'_2}
(\zeta_1,\zeta_2,\zeta_3)\,d\Omega\,=\frac{8\pi^2}{2J_1+1}\delta_{J_1J_2}
\delta_{M_1M_2}\delta_{M_1'M'_2}
$$
and (\ref{25}) we get:
\begin{equation}
\label{D2} {\Phi}^{(\pm)J}_{njK}(u_\alpha,v_\alpha,\theta_\alpha)
=\Bigl(\frac{\rho_\alpha}{\rho_\beta}\Bigr)^{1/2}\,
\sum_{\bar{K}}d_{\bar{K}K}^J(\vartheta){\Phi}^{(\pm)J}_{nj\bar{K}}(u_\beta,v_\beta,\theta_\beta).
\end{equation}
Now  we are multiplying the equation (\ref{19}) on the $(out)$
asymptotic state (\ref{21})-(\ref{22}) and integrating it by
coordinates $v$ and $\theta$ after which in the limit
$u\rightarrow+\infty$ the scattering matrix elements may be found:
\begin{eqnarray}
\label{D3}
\lim_{u\rightarrow+\infty}\Bigl\langle{\Phi}^{(+)J}_{njK}
(u_\alpha,v_\alpha,\theta_\alpha)
{\Phi}_{n'j\,'K'\,}^{(out)J}(u_\beta,v_\beta,\theta_\beta)
\Bigr\rangle_{v\theta}=\qquad\qquad\qquad\qquad
\nonumber\\
 \sum_{\widetilde{n}\,\widetilde{j\,}\,\widetilde{K}}
 \emph{S}^{J}_{\widetilde{n}\widetilde{j\,}\widetilde{K}\,\leftarrow\,njK}
 \,\lim_{u\rightarrow+\infty}
\Bigl\langle{\Phi}^{(-)J}_{\widetilde{n}\widetilde{j}\widetilde{K}}(u_\beta,v_\beta,\theta_\beta)
{\Phi}_{n'j\,'K'}^{(out)J}(u_\beta,v_\beta,\theta_\beta)
\Bigr\rangle_{v\theta}\nonumber\\
=\sum_{\widetilde{n}\,\widetilde{j\,}\,\widetilde{K}}
\delta_{n'\widetilde{n}}\,\delta_{j\,'\widetilde{j}}\,\delta_{K'\widetilde{K}}
\emph{S}^{J}_{\widetilde{n}\widetilde{j\,}\widetilde{K}\,\leftarrow\,njK}
=\emph{S}^{J}_{n'j\,'K'\,\leftarrow\,njK}\,.
 \end{eqnarray}

 So we found a new  expression for the scattering $\emph{\textbf{S}}$-matrix
 elements between $(in)$
and $(out)$  asymptotic states which are defined on different
$\alpha$ and $\beta$ surfaces:
\begin{eqnarray}
\label{D4} \emph{S}^{J}_{n'j\,'K'\,\leftarrow\,njK}=
\lim_{u\rightarrow+\infty}\Bigl\langle{\Phi}^{(+)J}_{njK}
(u_\alpha,v_\alpha,\theta_\alpha)
{\Phi}_{n'j\,'K'}^{(out)J}(u_\beta,v_\beta,\theta_\beta)
\Bigr\rangle_{v\theta}.
 \end{eqnarray}

 Note that in this
approach coordinate of translational motion $u$ plays similar role
as a usual time in the standard scattering theory and  will be
called \emph{intrinsic time}.

Now it is important to find the  $\emph{\textbf{S}}$-matrix
elements form  in the one coordinates system  $\alpha$ or
correspondingly $\beta$.

Putting the expression for overall wavefanction (\ref{27}) in the
(\ref{D4}) we can find:
\begin{equation}
\label{D5}
\emph{S}^{J}_{n'j\,'K'\,\leftarrow\,njK}(E)=\lim_{u\to+\infty}
\sqrt{\frac{p^{+}_{n'j\,'}}{p^-_{n j}}}\Bigl<\sum_{\bar{K}}
d_{\bar{K}K}^J(\vartheta){\Phi}^{(+)J}_{nj\bar{K}} (u,v,\theta)
{\Phi}_{n'j\,'K'}^{(out)J}(u,v,\theta)\Bigr>_{v_\beta\theta_\beta}.
\end{equation}
Since  the angle $\vartheta$ between vectors
$\textbf{q}_{0\alpha}$ and $\textbf{q}_{0\beta}$ in the $(out)$
channel (when $u\to+\infty$) is limited to $\pi$ and taking into
account that:
\begin{equation}
\label{D6}
  d_{(\bar{K}-K)}^J(\pi)=(-1)^{\bar{K}-K}\delta_{\bar{K}K},
\end{equation}
it is easy to get the expression
 for $\emph{\textbf{S}}$-matrix elements:
\begin{equation}
\label{D7}
S^{J}_{n'j\,'K'\,\leftarrow\,njK}(E)=\sqrt{\frac{p^{+}_{n'j\,'}}{p^-_{n
j}}}\lim_{u\rightarrow+\infty}\Bigl\langle{\Phi}^{(+)J}_{njK}
(u,v,\theta) {\Phi}_{n'j\,'K'}^{(out)J}(u,v,\theta)
\Bigr\rangle_{v\theta}.
\end{equation}
Besides in the expression (\ref{D7}) the coefficient
$\sqrt{{p^{+}_{\bar{n}\bar{j}}}/{p^-_{nj}}}$ is described flux
normalization constant:
\begin{equation}
\label{D8}\sqrt{\frac{p^{+}_{n'j\,'}}{p^-_{nj}}}=
\lim_{u\to+\infty}\Bigl(\frac{\rho_\alpha}{\rho_\beta}\Bigr)^{1/2}.
\end{equation}
In another words we proved that it is possible to represent the
full and asymptotic wavefunctions in term of same NCC system
without matching between reactant and product channels. It is very
important result of developed theory.

 So,  the matrix elements
$S_{n'j\,'K'\,\leftarrow\,njK}(E)$ describe the amplitudes of
transition probabilities between the sets of quantum numbers
$(njK)$  and $(n'j\,'K')$ correspondingly in the $(in)$ and
$(out)$ channels  at fixed total energy $E$ and will be called
reactive $S$-matrix elements.

\section{Solution of classical oscillator problem}

Now we turn to investigation of equation (\ref{51}) with natural
boundary conditions (\ref{53}). This equation can be solved with
the help of hypergeometric functions. The solution of Eq.
(\ref{51}), which has the form of an incident wave with positive
frequency $\omega_{i}>0$ in the $(in)$ channel (when
$u\to-\infty$) for a model frequency (\ref{52})  may be written in
the form (see for example \cite{BirDev}):

\begin{eqnarray}
\label{E1}
\sigma_{in}(\tau)=\omega_{i}^{-1/2}\exp\Bigl\{i\omega_{+}\tau+
i\frac{\omega_{-}}{\lambda\,\,}\ln[2\cosh{(\lambda{\tau})}]\Bigr\}
\times\nonumber\\
_2F_1\Bigl[1-i\frac{\omega_{-}}{\lambda},\,\,\,
-i\frac{\omega_{-}}{\lambda};\,\,1+i\frac{\omega_{i}}{\lambda};\,
\frac{1}{2}\bigl(1+\tanh{(\lambda{\tau})}\bigr)\Bigr]\stackrel{\sim}{
_{\tau \rightarrow
-\infty}}\,\,\omega_i^{-1/2}\exp{\bigl(i\omega_i\tau}\bigr),
 \end{eqnarray}
where

\begin{equation}
\label{E2} \omega_{\pm}=\frac{1}{2}(\omega_f\pm\omega_i), \quad
\omega_{i,f}=A_0\mp{A_1}.
\end{equation}
Now we can write the solution of Eq. (\ref{51}) which satisfies
another asymptotic condition in the $(out)$ channel:

\begin{eqnarray}
\label{E3} \sigma_{out}(\tau)=\omega_{f}^{-1/2}
\exp\Bigl\{i\omega_{+}\tau+
i\frac{\omega_{-}}{\lambda\,\,}\ln[2\cosh{(\lambda{\tau})}]\Bigr\}
\times\nonumber\\
_2F_1\Bigl[1-\frac{\omega_{-}}{\lambda},\,\,\,
-i\frac{\omega_{-}}{\lambda};\,\,1-i\frac{\omega_{f}}{\lambda};\,
\frac{1}{2}\bigl(1-\tanh{(\lambda{\tau})}\bigr)\Bigr]\stackrel{\sim}{
_{\tau \rightarrow
+\infty}}\,\,\omega_f^{-1/2}\exp{\bigl(i\omega_f\tau}\bigr).
\end{eqnarray}
Both these solutions are connected with each other by Bogoliubov
\cite{Bog} linear transformations:

\begin{eqnarray}
\label{E4}
\sigma_{in}(\tau)=C_1{\sigma}_{out}(\tau)-C_2{\sigma}_{out}^{\ast}(\tau),
\end{eqnarray}
Coefficients $C_1$ and $C_2$ are easily calculated, using of
properties of hypergeometric functions:

\begin{eqnarray}
\label{E5}
 C_1=<\sigma_{in}(\tau)\sigma_{out}{(\tau})>_{\tau}=
 \Bigl(\frac{{\omega_f}}{\omega_i}\Bigr)^{1/2}
\frac{\Gamma(1+i\omega_i/\lambda)\Gamma(i\omega_f/\lambda)}
{\Gamma(1+i\omega_+/\lambda)\Gamma(i\omega_{-}/\lambda)},
\nonumber\\
 C_2=-<\sigma_{in}(\tau)\sigma_{out}^{\ast}{(\tau})>_{\tau}=
 \Bigl(\frac{{\omega_f}}{\omega_i}\Bigr)^{1/2}
\frac{\Gamma(1+i\omega_i/\lambda)\Gamma(-i\omega_f/\lambda)}
{\Gamma(1-i\omega_+/\lambda)\Gamma(-i\omega_{-}/\lambda)}.
\end{eqnarray}
Recall that similar to (\ref{E4}), the following expression may be
written for the solution ${\sigma}_{out}(\tau)$:

\begin{eqnarray}
\label{E6}
\sigma_{out}(\tau)=C_1{\sigma}_{in}(\tau)-C_2{\sigma}_{in}^{\ast}(\tau),
\end{eqnarray}
Now taking into account (\ref{E4}), (\ref{E5}) and (\ref{E6}) for
coefficient $C_1$ and $C_2$ we obtain:

\begin{eqnarray}
\label{E7}
|C_1|^2=\frac{\sinh^2(\pi\omega_+/\lambda)}{\sinh(\pi\omega_{in})
\sinh(\pi\omega_{out})},\quad
|C_2|^2=\frac{\sinh^2(\pi\omega_-/\lambda)}{\sinh(\pi\omega_{in})
\sinh(\pi\omega_{out})}.
\end{eqnarray}
Finally from (\ref{E7}) we can find the normalization expression:
\begin{eqnarray}
\label{E8} |C_1|^2- |C_2|^2=1.
\end{eqnarray}
So, taking into account (\ref{E3}) and (\ref{E7}), we get the
exact solution of Eq. (\ref{51}), which we must use for future
analytical computations, in the form (\ref{E4}).

\end{document}